\documentclass[draft,tightenlines,nofootinbib,preprint,aps,eqsecnum]{revtex4}

\usepackage{latexsym}

\newcommand{\beq}{\begin{equation}}
\newcommand{\eeq}{\end{equation}}
\newcommand{\bea}{\begin{eqnarray}}
\newcommand{\eea}{\end{eqnarray}}
\newcommand{\cir}{{\buildrel \circ \over =}}
\newcommand{\sgn}{\epsilon}
\newcommand{\eo}{{}^4{\buildrel \circ \over E}}

\begin{document}

\title{Dirac-Bergmann Constraints in Relativistic Physics: Non-Inertial Frames,  Point Particles, Fields  and Gravity }

\medskip

\author{Luca Lusanna}

\affiliation{ Sezione INFN di Firenze\\ Polo Scientifico\\ Via Sansone 1\\
50019 Sesto Fiorentino (FI), Italy\\  E-mail: lusanna@fi.infn.it}

\today

\begin{abstract}

There is a review of the physical theories needing Dirac-Bergmann theory of constraints at the Hamiltonian level due to the existence of gauge symmetries.
It contains:
\medskip

i) the treatment of systems of point particles in special relativity both in inertial and non-inertial frames with a Wigner-covariant way of eliminating 
relative times in relativistic bound states;

\medskip

ii) the description of the electro-magnetic field in relativistic atomic physics and of Yang-Mills fields in absence of Gribov ambiguity in particle physics;

\medskip

iii) the identification of the inertial gauge variables and of the physical variables in canonical ADM tetrad gravity in 
presence of the electro-magnetic field and of charged scalar point particles in asymptotically Minkowskian space-times 
without super-translations by means of a Shanmugadhasan canonical transformation to a York canonical basis adapted to ten 
of the 14 first-class constraints and the definition of the Hamiltonian Post-Minkowskian weak field limit.

\bigskip

Review paper for a chapter of a future book.

\end{abstract}

\bigskip

\maketitle

\section{Introduction}

Most of the relevant interactions in physics are described by singular Lagrangians implying the presence
of Dirac-Bergmann constraints \cite{1,2,3} at the Hamiltonian level, whose treatment was given in a previous
chapter. This happens for electro-magnetism, for the standard model of particle physics ($SU(3) \times SU(2) \times U(1)$
Yang-Mills fields) and its extensions, for Einstein theory of gravity and for all its generally covariant variants.
Also the description of relativistic classical and quantum point particles, needed for bound states in the particle
approximation of quantum field theory (QFT), requires Hamiltonian constraints for the elimination of relative times
(no time-like excitation is seen in spectroscopy).
In all these theories the main problem at the classical level is the identification of the gauge-invariant physical
degrees of freedom, the so called Dirac observables (DO). Instead the main open problem at the quantum level is whether
one has to quantize only the DO's or also the gauge variables shifting the search of the physical observables after
quantization like in the BRST approach.

\bigskip

I will present a review of the main properties of relativistic constrained systems (from particles to gauge theories) at the classical level
first in inertial and non-inertial frames in special relativity (SR) and then in the dynamical space-times
of Einstein general relativity (GR),  based on my personal viewpoint,
with some comments on the weak points of the existing quantization approaches.
\medskip

Besides Dirac's book \cite{1} and Ref.\cite{4} I recommend the books in Refs.\cite{5,6} for an extended treatment of many aspects of the theory also at the quantum level
(included the BRST approach). Other books on the subject are in Refs. \cite{7,8,9,10}. Instead there is no good treatment of constrained
systems in mathematical physics and differential geometry: there are only partial treatments for finite-dimensional systems like presymplectic geometry \cite{11,12}
(see Refs.\cite{13,14} and their bibliography for recent contributions) without any extension to infinite-dimensional systems like field theory \cite{15}.

\bigskip

In Section II I show the importance of constraint theory in special relativity for the description of isolated systems in non-inertial and inertial frames in
Minkowski space-time by means of parametrized Minkowski theories and for the development of  classical and quantum relativistic mechanics of point particles.

In Section III I give the treatment of the electro-magnetic field needed to formulate relativistic atomic physics. Then,
after reviewing the problems of gauge symmetries some results on Yang-Mills fields in absence of Gribov
ambiguity are described.

In Section IV there is the treatment of Einstein gravity with constraint theory in a family of space-times reducing to Minkowski space-time when the Newton constant is
switched off: in these asymptotically Minkowskian space-time there is the asymptotic ADM Poincare' group needed for the inclusion of the standard particle model in
the matter. I describe the Hamiltonian constraints of ADM tetrad gravity (needed for the inclusion of fermions) after a 3+1 splitting of the space-time
like in SR and the Shanmugadhasan canonical transformation to the York canonical basis adapted to 10 of the 14 first-class constraints.  I make some
comments on the gauge variables (like the York time), on the search of DO's and on the canonical quantization of gravity. Then I give an idea of relativistic atomic physics
plus gravity, of its Hamiltonian Post-Minkowskian (HPM) and Post-Newtonian (PN) approximations and of the possible role of the York time gauge variable in reducing at least part of dark matter to a choice of conventions in relativistic metrology.

Some open problems are described in the Conclusions.

\section{Non-Inertial Frames in Special Relativity and Parametrized Minkowski Theories}

In this Section after the definition of non-inertial frames in Minkowski space-time (Subsection A), of the inertial rest frame
and of the existing notions of relativistic center of mass of isolated systems (Subsection B) there is
the definition of parametrized Minkowski theories for isolated systems (Subsection C) and of relativistic
quantum mechanics for point particles (Subsection D).

\subsection{Non-Inertial Frames in Minkowski Spacetime}

Assume that the world-line $x^{\mu}(\tau)$ of an arbitrary time-like
observer carrying a standard atomic clock is given either in Minkowski space-time
or in the quoted class of Einstein space-times: $\tau$ is an arbitrary
monotonically increasing function of the proper time of this clock. Then one
gives an admissible 3+1 splitting of the asymptotically flat space-time, namely a nice
foliation with space-like instantaneous 3-spaces $\Sigma_{\tau}$. It is the
mathematical idealization of a protocol for clock synchronization: all the
clocks in the points of $\Sigma_{\tau}$ sign the same time of the atomic
clock of the observer (see Ref.\cite{16} for a review on relativistic metrology). The observer and the
foliation define a {\it global non-inertial reference frame} after a choice of
4-coordinates. On each 3-space $\Sigma_{\tau}$  one chooses curvilinear
3-coordinates $\sigma^r$ having the observer as origin.
The quantities $\sigma^A = (\tau; \sigma^r)$ are the either Lorentz- or world-scalar and observer-dependent
\textit{radar 4-coordinates}, first introduced by Bondi \cite{17,18}. \medskip

If $x^{\mu} \mapsto \sigma^A(x)$ is the coordinate transformation from
world 4-coordinates $x^{\mu}$ having the observer as origin to radar 4-coordinates \cite{19,20}, its inverse $\sigma^A
\mapsto x^{\mu} = z^{\mu}(\tau ,\sigma^r)$ defines the \textit{embedding}
functions $z^{\mu}(\tau ,\sigma^r)$ describing the 3-spaces $\Sigma_{\tau}$
as embedded 3-manifolds into the asymptotically flat space-time.
Let $z^{\mu}_A(\tau, \sigma^u) = \partial\, z^{\mu}(\tau, \sigma^u) / \partial\, \sigma^A$ denote the gradients of the
embedding functions with respect to the radar 4-coordinates.
The space-like 4-vectors $z^{\mu}_r(\tau ,\sigma^u)$ are tangent to $\Sigma_{\tau}$, so that the unit
time-like normal $l^{\mu}(\tau ,\sigma^u)$ is proportional to $\epsilon^{\mu}{}_{%
\alpha \beta\gamma}\, [z^{\alpha}_1\, z^{\beta}_2\, z^{\gamma}_3](\tau
,\sigma^u)$ ($\epsilon_{\mu\alpha\beta\gamma}$ is the Levi-Civita tensor).
Instead $z^{\mu}_{\tau}(\tau, \sigma^u)$ is a time-like 4-vector skew with respect to the 3-spaces leaves of the foliation.
In SR one has $z^{\mu}_{\tau}(\tau ,\sigma^r) = [N\, l^{\mu} + N^r\,
z^{\mu}_r](\tau ,\sigma^r)$ with $N(\tau ,\sigma^r) = \epsilon\,
[z^{\mu}_{\tau}\, l_{\mu}](\tau ,\sigma^r) = 1 + n(\tau, \sigma^r) > 0$ and $
N_r(\tau ,\sigma^r) = - \epsilon\, [z^{\mu}_{\tau}\, \eta_{\mu\nu}\, z_r^{\mu}](\tau ,\sigma^r)$ being the
lapse and shift functions respectively of the {\it global non-inertial frame} of Minkowski space-time so defined.\medskip

\bigskip

The induced 4-metric ${}^4g_{AB}(\tau, \sigma^r) = z^{\mu}_A(\tau, \sigma^r)\,
z^{\nu}_B(\tau, \sigma^r)\, {}^4\eta_{\mu\nu}$ has signature $\sgn\, (+---)$
like the flat Minkowski metric ${}^4\eta_{\mu\nu}$  with $\sgn =
\pm$ (the particle physics, $\sgn = +$, and general relativity,
$\sgn = -$, conventions). From now on we shall denote
the curvilinear 3-coordinates $\sigma^r$ with the notation $\vec
\sigma$ for the sake of simplicity. Usually the convention of sum
over repeated indices is used, except when there are too many
summations. The symbol $\approx$ means Dirac weak equality, while the symbol
$\cir$ means evaluated by using the equations of motion.

\subsection{The Wigner-Covariant Rest-Frame Instant Form of Dynamics and the Relativistic Center of Mass for Isolated Systems}

In SR we can restrict ourselves to inertial frames
and define the {\it inertial rest-frame instant form of dynamics for
isolated systems} by choosing the 3+1 splitting corresponding to the
intrinsic inertial rest frame of the isolated system centered on an
inertial observer: the instantaneous Euclidean 3-spaces, named Wigner 3-space
due to the fact that the 3-vectors inside them are Wigner spin-1
3-vectors \cite{19}, are orthogonal to the conserved 4-momentum
$P^{\mu}$ of the configuration. In Ref.\cite{19} there is also the extension to admissible {\it
non-inertial rest frames}, where $P^{\mu}$ is orthogonal to the
asymptotic space-like hyper-planes to which the instantaneous non-Euclidean
3-spaces tend at spatial infinity.

\bigskip

The simplest form of the embedding of the Wigner 3-spaces in
Minkowski space-time described in the inertial frame of an arbitrary
inertial observer is\medskip

\beq
 z_W^{\mu}(\tau, \vec \sigma) =  Y^{\mu}(\tau) +
 \epsilon^{\mu}_r(\vec h)\, \sigma^r = Y^{\mu}(0) +
 \Lambda^{\mu}{}_A(\vec h)\, \sigma^A,
 \label{2.3}
 \eeq

\noindent where $Y^{\mu}(\tau ) = Y^{\mu}(0) + h^{\mu}\, \tau  =
z^{\mu}_W(\tau ,\vec 0)$ is the world-line of the external
Fokker-Pryce 4-center of inertia.
The Lorentz matrix $\Lambda^{\mu}{}_A(\vec h)$ is obtained from the standard Wigner
boost $\Lambda^{\mu}{}_{\nu}(P^{\alpha}/Mc)$, sending the time-like
4-vector $P^{\mu}/Mc$ into $(1; 0)$, by transforming the index $\nu$
into an index adapted to radar 4-coordinates ($\Lambda^{\mu}{}_{\nu}
\mapsto \Lambda^{\mu}{}_A$). One has $\epsilon^{\mu}_{\tau}(\vec h) = {{P^{\mu}}\over {M c}} = u^{\mu}(P)
= h^{\mu} = \Big(\sqrt{1 + {\vec h}^2}; \vec h\Big) =
\Lambda^{\mu}{}_{\tau}(\vec h)$, $\epsilon^{\mu}_r(\vec h) = \Big( h_r; \delta^i_r + {{h^i\, h_r}\over
 {1 + \sqrt{1 + {\vec h}^2}}}\Big) = \Lambda^{\mu}{}_r(\vec h)$  with ${}^4\eta_{\mu\nu}\,
\epsilon^{\mu}_A(\vec h)\, \epsilon^{\nu}_B(\vec h) = {}^4\eta_{AB}$ and
$\sgn\, P^2 = M^2\, c^2$.

\bigskip

The form of this embedding is a consequence of the clarification of
the notion of relativistic center of mass of an isolated system
after a century of research \cite{21,22}. It turns out that there are only three
notions of collective variables, which can be built using {\it
only} the Poincar\'e generators (they are {\it non-local} quantities
expressed as integrals over the whole 3-space $\Sigma_{\tau}$) of the
isolated system: the {\it canonical non-covariant Newton-Wigner 4-center
of mass} (or center of spin) ${\tilde x}^{\mu}(\tau)$, the
{\it non-canonical covariant Fokker-Pryce 4-center of inertia}
$Y^{\mu}(\tau)$ and the {\it non-canonical non-covariant M$\o$ller
4-center of energy} $R^{\mu}(\tau)$. All of them tend to the
Newtonian center of mass in the non-relativistic limit.
These three variables can be expressed as known
functions:
a) of the Lorentz-scalar rest time $\tau = c\, T_s = h \cdot \tilde
x = h \cdot Y = h \cdot R$;
b) of canonically conjugate Jacobi data (frozen Cauchy data) $\vec h
= \vec P/Mc$ and $\vec z = Mc\, {\vec x}_{NW}(0)$ ($\{ z^i, h^j\} =
\delta^{ij}$) \footnote{The 3-vector ${\vec x}_{NW}(\tau )$ is the
standard Newton-Wigner non-covariant 3-position, classical
counterpart of the corresponding position operator; the use of $\vec
z$ avoids taking into account the mass spectrum of the isolated
system in the description of the center of mass.};
c) of the invariant mass $Mc = \sqrt{\sgn\, P^2}$ of the isolated
system;
d) of the rest spin ${\vec S}$ of the isolated system.

\bigskip

In each Lorentz frame one has different pseudo-world-lines
describing $R^{\mu}$ and ${\tilde x}^{\mu}$: the canonical 4-center
of mass ${\tilde x}^{\mu}$ {\it lies in between} $Y^{\mu}$ and
$R^{\mu}$ in every frame. This leads to the existence of
the {\it M$\o$ller non-covariance world-tube} \cite{23}, around the world-line
$Y^{\mu}$ of the covariant non-canonical Fokker-Pryce 4-center of
inertia $Y^{\mu}$. The {\it invariant radius} of the tube is $\rho
=\sqrt{- W^2}/p^2 = |\vec S|/\sqrt{P^2}$ where ($W^2 = - P^2\, {\vec
S}^2$ is the Pauli-Lubanski invariant when $P^2 > 0$). This
classical intrinsic radius delimits the non-covariance effects (the
pseudo-world-lines) of the canonical 4-center of mass ${\tilde
x}^{\mu}$. They are not detectable because the M$\o$ller radius is of the order of the
Compton wave-length: an attempt to test its interior would mean to
enter in the quantum regime of pair production.

\bigskip

Every isolated system  can be visualized \cite{19} as
a decoupled {\it non-covariant collective (non-local) pseudo-particle}
described by the frozen Jacobi data $\vec z$, $\vec h$ carrying a
{\it pole-dipole structure}, namely the invariant mass $M\, c$ and
the rest spin ${\vec S}$ of the system, and with an associated {\it
external} realization  of the Poincar\'e group (the last term in the
Lorentz boosts induces the Wigner rotation of the 3-vectors inside
the Wigner 3-spaces): $P^{\mu} = M\, c\, h^{\mu} = M\, c\, \Big(\sqrt{1 + {\vec
h}^2}; \vec h\Big)$, $J^{ij} = z^i\, h^j - z^j\, h^i + \epsilon^{ijk}\, S^k$, $K^i =
J^{oi} = - \sqrt{1 + {\vec h}^2}\, z^i + {{(\vec S \times \vec
h)^i}\over {1 + \sqrt{1 + {\vec h}^2}}}$,  satisfying the Poincar\'e algebra.
The universal breaking of Lorentz covariance  connected to these
decoupled non-local collective variables  is irrelevant because
all the dynamics of the isolated system lives inside the Wigner
3-spaces and is Wigner-covariant \footnote{The use of $\vec{z}$ avoids taking into account the mass spectrum of the
isolated system at the quantum kinematical level as already said. Moreover,
since the center of mass is decoupled, its non-covariance is irrelevant: like for the
wave function of the universe, who will observe it?}. Inside the Wigner 3-spaces the
system is described by an internal 3-center of mass with a conjugate
3-momentum and by relative variables and there is an {\it unfaithful internal}
realization of the Poincar\'e group \cite{19} (whose generators are determined
by using the energy-momentum tensor $T^{\mu\nu}$ of the isolated system):
 $Mc =  \int d^3\sigma\, T^{\tau\tau}(\tau , \vec \sigma)$,
  ${\bar S}^r = {1\over 2}\, \delta^{rs}\, \epsilon_{suv}\, \int d^3\sigma\,
  \sigma^u\, T^{\tau v}(\tau , \vec \sigma)$,
 ${\cal P}^r = \int d^3\sigma\, T^{\tau r}(\tau , \vec \sigma)
 \approx 0$, ${\cal K}^r = - \int d^3\sigma\, \sigma^r\,
 T^{\tau\tau}(\tau ,\vec \sigma) \approx 0$.
The internal 3-momentum, conjugate to the internal 3-center of mass,
vanishes due the rest-frame condition
\footnote{Due to the rest-frame condition ${\vec {\cal
P}} \approx 0$, we have ${\vec q}_+ \approx {\vec
R}_+ \approx {\vec y}_+$, where ${\vec q}_+$ is the internal
canonical 3-center of mass (the internal Newton-Wigner position),
${\vec y}_+$ is the internal Fokker-Pryce 3-center of inertia and
${\vec R}_+$ is the internal M$\o$ller 3-center of energy. As a
consequence there is a unique internal 3-center of mass, which is
eliminated by the vanishing of the internal Lorentz boosts ${\cal K}^r \approx 0$.}.
To avoid a double counting of
the center of mass, i.e. an external one and an internal one, the
internal (interaction-dependent)  Lorentz boosts must also vanish.
The only non-zero internal generators are the invariant mass $M\, c$
and the rest spin ${\vec S}$ and the dynamics is re-expressed only
in terms of {\it internal Wigner-covariant relative variables}.
In the Wigner 3-spaces  the effective Hamiltonian is the invariant mass of the
isolated system: $H = M\, c$.

 \bigskip

As shown in Refs.\cite{19,24} for N free positive energy spinless particles their world-lines  are
parametrized in terms of Wigner 3-vectors ${\vec \eta}_i(\tau)$, $i
= 1,..,N$, in the following way $x^{\mu}_i(\tau) = z^{\mu}(\tau, {\vec \eta}_i(\tau))$:
one eliminates the possibility to have
time-like excitations in the spectrum of relativistic bound states, because
inside each 3-space only space-like correlations among the particles
are possible.
At the Hamiltonian level the basic canonical variables describing
the particle are ${\vec \eta}_i(\tau)$ and their canonically
conjugate momenta ${\vec \kappa}_i(\tau)$: $\{ \eta^r_i(\tau),
\kappa^s_j(\tau)\} = \delta_{ij}\, \delta^{rs}$. The standard
momenta of the positive-energy scalar particles are $p_i^{\mu}(\tau) = \Lambda^{\mu}{}_A(\vec h)\,
\kappa^A_i(\tau)$, $\kappa^A_i(\tau) = ( E_i(\tau); \kappa_{ir}(\tau) )$.
For free particles we have $E_i(\tau) = \sqrt{m_i^2\, c^2
+ {\vec \kappa}_i^2(\tau)}$, $\sgn\, p_i^2 = m_i^2\, c^2$ and $M c =
\sum_i\, E_i$.
The internal generators have the following expression
in the rest frame $M\, c = {1\over c}\, {\cal E}_{(int)} = \sum_{i=1}^N\,
\sqrt{m_i^2\, c^2 + {\vec \kappa}^2_i}$,
${\vec {\cal P}} = \sum_{i=1}^2\, {\vec \kappa}_i \approx
0$, $\vec S = {\vec {\cal J}} = \sum_{i=1}^2\,
{\vec \eta}_i \times {\vec \kappa}_i$,
${\vec {\cal K}} = - \sum_{i=1}^2\, {\vec \eta}_i\, \sqrt{m_i^2\,
c^2 + {\vec \kappa}_i^2} \approx 0$.

\medskip

In the interacting case it is $E_i \not= \sqrt{m_i^2\, c^2 + {\vec
\kappa}_i^2(\tau)}$ and $\sgn\, p_i^2 \not= m_i^2\, c^2$. Instead
$E_i(\tau)$ must be deduced from the form of the invariant mass $M
c$, which is the Hamiltonian for the $\tau$-evolution in the Wigner
3-spaces. This description is not in contrast with scattering theory, where
$\sgn\, p_i^2 = m_i^2\, c^2$  holds asymptotically for the {\it in}
and {\it out} free particles (no interpolating description due to
Haag no-go theorem for the interaction picture: see Ref.\cite{24} for
the way out, at least at the classical level, from this theorem in
the 3+1 point of view), if the action-at-a-distance potentials (and
also interactions with electro-magnetic fields) go to zero for large
separations of the particles inside Wigner 3-spaces (the {\it
cluster separability} in action-at-a-distance theories).
See Refs.\cite{25,26,27,28} for the description of relativistic bound states.
In relativistic kinetic theory and in relativistic statistical
mechanics $E_i = \sqrt{m_i^2\, c^2 + {\vec \kappa}_i^2(\tau)}$ holds
for a gas of non-interacting particles, otherwise it has to be
replaced with an expression dictated by the type of the existing
interactions (see Ref.\cite{29} on the relativistic microcanonical ensemble).

\bigskip

In the two-body case,
by introducing the notation ${\vec \eta}_+$, ${\vec \kappa}_+ =
{\vec {\cal P}}$, with a canonical transformation we get the
following internal collective and relative variables
 ${\vec \eta}_+ = {\frac{{m_1}}{m}}\, {\vec \eta}_1 +
{\frac{{m_2}}{m}}\, {\vec\eta}_2$,
${\vec \rho} = {\vec \eta}_1 - {\vec \eta}_2$,
${\vec \kappa}_+ = {\vec \kappa}_1 + {\vec \kappa}_2 \approx 0$,
 ${\vec \pi} =  {\frac{{m_2}}{m}}\, {\vec \kappa}_1 - {\
\frac{{m_1}}{m}}\, {\vec \kappa}_2$.
The collective variable ${\vec \eta}_+(\tau )$ has to be determined
in terms of ${\vec \rho}(\tau )$ and ${\vec \pi}(\tau )$ by means of
the gauge fixings ${\vec {\mathcal{K}}}\,  \approx 0$. For two {\it free} particles
we get ${\vec \eta}_+(\tau) \approx {\vec \eta}(\tau) =
 {{{{m_1}\over m}\, \sqrt{m_2^2\,
c^2 + {\vec \pi}^2(\tau)} - {{m_2}\over m}\, \sqrt{m_1^2\, c^2 +
{\vec \pi}^2(\tau)}}\over {\sqrt{m_1^2\, c^2 + {\vec \pi}^2(\tau)} +
\sqrt{m_2^2\, c^2 + {\vec \pi}^2(\tau)}}}\, {\vec \rho}(\tau)$
(${\vec \eta}_+(\tau) \approx 0$ for $m_1 = m_2$).
In the interacting case the rest-frame conditions ${\vec \kappa}_+
\approx 0$ and  the conditions eliminating the internal 3-center of
mass ${\vec {\cal K}} \approx 0$ will determine ${\vec \eta}_+$ in
terms of the relative variables ${\vec \rho}$, ${\vec \pi}$ in an
interaction-dependent way.
Then the relative variables satisfy Hamilton equations with the
invariant mass $M({\vec \rho}, {\vec \pi})$ as Hamiltonian and the
particle world-lines $x^{\mu}_i(\tau )$ can be rebuilt \cite{28}
(they are 4-vectors but {\it not canonical variables} like in most of the approaches,
with this non-commutative structure induced by the Lorentz signature of the space-time).

\subsection{Parametrized Minkowski Theories for Isolated Systems}

In the global non-inertial frames of Minkowski space-time it is
possible to describe isolated systems (particles, strings, fields,
fluids) admitting a Lagrangian formulation  by means of {\it
parametrized Minkowski theories} \cite{19,30} (see Refs.\cite{31} for reviews).
The existence of a Lagrangian, which can be coupled  to an external
gravitational field, makes possible the determination of the matter
energy-momentum tensor $T^{\mu\nu}$ and of the ten conserved Poincar\'e
generators $P^{\mu}$ and $J^{\mu\nu}$ (assumed finite) of every
configuration of the isolated system.
\bigskip

First of all one must replace the matter variables of the isolated
system with new ones knowing the clock synchronization convention
defining the 3-spaces $\Sigma_{\tau}$.
As said a positive (or negative-) energy relativistic particle with world-line
$x^{\mu}(\tau ) = z^{\mu}(\tau , \vec \eta(\tau ))$ is described by the 3-coordinates
$\eta^r(\tau )$ defined by the intersection of its world-line with
$\Sigma_{\tau}$. A Klein-Gordon
field $\tilde \phi (x)$ will be replaced with $\phi(\tau , \vec \sigma)
= \tilde \phi (z(\tau , \vec \sigma))$; the same for every other field.
Then one replaces the external gravitational 4-metric in the coupled
Lagrangian with the 4-metric ${}^4g_{AB}(\tau , \vec \sigma) = z^{\mu}_A(\tau, \vec \sigma)\,
z^{\nu}_B(\tau, \vec \sigma)\, {}^4\eta_{\mu\nu}$, which is
a functional of the embedding
\footnote{As an example one may consider N free scalar particles with masses
$m_i$, sign of the energy $\eta_i = \pm$ and world-lines
$x^{\mu}_i(\tau) = z^{\mu}(\tau, {\vec \eta}_i(\tau))$, $i=1,..,N$.
In parametrized Minkowski theories they are described by the following
action depending on the configurational variables $\eta^r_i(\tau)$
and $z^{\mu}(\tau, \vec \sigma)$:
$S =\int d\tau\, d^{3}\sigma \,{\cal L}(\tau , \vec \sigma) = \int d\tau\, d^3\sigma\, \Big(
 -\sum_{i=1}^{N}\, \delta^3(\sigma^u - \eta^u_i(\tau ))
 \, m_ic\, \eta_i\, \sqrt{\sgn\, [{}^4g_{\tau \tau }(\tau ,\sigma^u) + 2\,
 {}^4g_{\tau r}(\tau ,\sigma^u)\, {\dot{\eta}}_i^r(\tau ) + {}^4g_{rs }(\tau
 ,\sigma^u)\, {\dot{\eta}}_i^r(\tau )\, {\dot{\eta}}_i^s(\tau
 )]} \Big)$.}.

\bigskip

Parametrized Minkowski theories are defined by the resulting
Lagrangian depending on the given matter and on the embedding
$z^{\mu}(\tau , \vec \sigma)$. The resulting action is invariant under
the {\it frame-preserving diffeomorphisms} $\tau \mapsto \tau^{'}(\tau, \sigma^u)$, $\sigma^r
\mapsto \sigma^{' r}(\sigma^u)$ firstly introduced in Ref.\cite{32}.
As a consequence, there are four first-class constraints with
exactly vanishing Poisson brackets (an Abelianized analogue of the
super-Hamiltonian and super-momentum constraints of canonical
gravity) determining the momenta $\rho_{\mu}(\tau, \vec \sigma)$ conjugated to the embeddings in
terms of the matter energy-momentum tensor:
$\rho_{\mu}(\tau, \vec \sigma) - \sqrt{\gamma(\tau,
\vec \sigma)}\, \Big[l_{\mu}\, T_{\perp\perp} - z_{r\mu}\, h^{rs}\,
T_{\perp s}\Big](\tau, \vec \sigma) \approx 0$, where
$h^{rs}$ is the inverse of the induced 3-metric on the 3-spaces,
$T_{\perp\perp} = l_{\mu}\, l_{\nu}\, T^{\mu\nu}$
and $ T_{\perp r} = l_{\mu}\, z_{r\nu}\, T^{\mu\nu}$.
The ten external Poincar\'e generators are
 $P^{\mu} = \int d^3\sigma\, \rho^{\mu}(\tau, \vec \sigma)$,
 $J^{\mu\nu} = \int d^3\sigma\, \Big(z^{\mu}\, \rho^{\nu} -
 z^{\nu}\, \rho^{\mu}\Big)(\tau, \vec \sigma)$.
This implies that the
embeddings $z^{\mu}(\tau ,\sigma^r)$ are {\it gauge variables}, so
that {\it all the admissible non-inertial or inertial frames are
gauge equivalent}, namely physics does {\it not} depend on the clock
synchronization convention and on the choice of the 3-coordinates
$\sigma^r$: only the appearances of phenomena change by changing the
notion of instantaneous 3-space \footnote{In
Refs.\cite{33,34} there is the definition of {\it parametrized Galilei
theories}, non relativistic limit of the parametrized Minkowski
theories. Also the inertial and non-inertial frames in Galilei
space-time are gauge equivalent in this formulation.}.
To describe the physics in a given admissible non-inertial frame
described by an embedding $z^{\mu}_F(\tau, \sigma^u)$ one must add
the gauge-fixings $z^{\mu}(\tau, \sigma^u) - z^{\mu}_F(\tau,
\sigma^u) \approx 0$.

\bigskip

The same description can be given for spinning particles \cite{35}
\footnote{The pseudo-classical description of the spin is made with
Grassmann variables, whose quantization leads to Clifford algebras including
Pauli and Dirac matrices.}, massless particles \cite{36} and 2-level atoms \cite{37}.
The open and closed Nambu string have been studied \cite{38,39}
in the stratum $\sgn\, p^2 > 0$: Abelian Lorentz
scalar constraints and gauge variables have been found and
globally decoupled, and a redundant set of DO's has been found.
Then the rest-frame formulation and a basis of DO's has been found for
the open Nambu string \cite{40}.

\subsection{Quantum Relativistic Point Particles for Bound States}

A new formulation of {\it relativistic quantum mechanics} (RQM) in the
Wigner 3-spaces of the inertial rest frame  is developed in
Ref.\cite{41} in absence of the electro-magnetic field. It includes
all the known results about relativistic bound states (absence of
relative times) and avoids the causality problems of the Hegerfeldt
theorem \cite{42} (the instantaneous spreading of wave packets). See the
bibliography of Ref. \cite{41} for all the previous non satisfactory attempts to
define a consistent RQM.
\medskip

In non-relativistic quantum mechanics (NRQM) the Hilbert space of
a quantum two-body system can be described in the three following
{\it unitarily equivalent} ways \cite{41}: A) as the
tensor product $H=H_{1}\otimes H_{2}$, where $H_{i}$ are the Hilbert spaces
of the two particles (separability of the two subsystems as the zeroth
postulate of NRQM); B) as the tensor product $H=H_{com}\otimes H_{rel}$,
where $H_{com}$ is the Hilbert space of the decoupled free Newton center of
mass and $H_{rel}$ the Hilbert space of the relative motion (in the
interacting case only this presentation implies the separation of variables
in the Schroedinger equation); C) as the tensor product $H=H_{HJcom}\otimes
H_{rel}$, where $H_{HJcom}$ is the Hilbert space of the frozen Jacobi data
of the Newton center of mass (use is made of the Hamilton-Jacobi
transformation). Each of these three presentations gives rise to a different notion of
entanglement due to the different notion of separable subsystems.

\medskip

As shown in Ref.\cite{41}, at the relativistic level the elimination of the
relative times of the particles  and the treatment of the relativistic collective
variables allows \textit{only the presentation C)}, i.e. $H=H_{HJcom}\otimes
H_{rel}$ with $H_{HJcom}$ being the Hilbert space associated to the
quantization of the canonically conjugate frozen Jacobi data $\vec{z}$ and $%
\vec{h}$  and $H_{rel}$ is the Hilbert space of the Wigner-covariant relative
3-coordinates and 3-momenta.
As a consequence, at the relativistic level the zeroth postulate of
non-relativistic quantum mechanics does not hold: the Hilbert space
of composite systems is not the tensor product of the Hilbert spaces
of the sub-systems. Contrary to the standard notion of separability
(separate objects have their independent real states) one
gets a {\it kinematical spatial non-separability} induced by the
need of clock synchronization for eliminating the relative times
in relativistic bound states \footnote{If one considers the tensor product $H_{1}\otimes H_{2}$ of two massive
Klein-Gordon particles most of the states will have one particle allowed to
be the absolute future of the other due to the lack of restrictions on the
relative times. Only in S-matrix theory is this irrelevant since one takes
the limit for infinite future and past times. Therefore this is a unitarily
inequivalent quantization.} and
to be able to formulate a well-posed relativistic Cauchy problem.
Moreover one has the {\it non-locality} of the non-covariant
external center of mass which implies its {\it non-measurability}
with local instruments. While its conjugate momentum operator must be well
defined and self-adjoint, because its eigenvalues describe the
possible values for the total momentum of the isolated system (the
momentum basis is therefore a {\it preferred basis} in the Hilbert
space), it is not clear whether it is meaningful to define
center-of-mass wave packets.
These {\it non-locality} and {\it kinematical spatial
non-separability} are due to the Lorentz signature of Minkowski
space-time.

 \medskip

Let us remark that instead of starting from the physical Hilbert space
containing the frozen Jacobi data, one could first define an un-physical
Hilbert space containing the Jacobi data and the 3-position and 3-momenta of
the particles (in it we have the same kind of separability as in the
presentation A) of NRQM) and then define the physical Hilbert space by
imposing the rest-frame conditions at the quantum level with the
Gupta-Bleuler method. However there is the risk to get an inequivalent
quantum theory due to the complex form of the internal boosts.

\medskip

The quantization defined in Ref.\cite{41} leads to a first
formulation of a theory for {\it relativistic entanglement}, which
is deeply different from the non-relativistic entanglement due to
the kinematical non-locality and spatial non-separability. To have
control on the Poincar\'e group one needs an isolated systems
containing all the relevant entities (for instance both Alice and
Bob) of the experiment under investigation and also the environment
when needed. One has to learn to reason in terms of relative
variables adapted to the experiment like molecular physicists do
when they look to the best system of Jacobi coordinates adapted to
the main chemical bonds in the given molecule. This theory has still
to be developed together with its extension to non-inertial rest
frames. See Refs.\cite{43,44} for more details on the notion of relativistic entanglement
and on the problem of localization of particles.

\section{Classical Gauge Fields in the Rest-Frame Instant Form}

By means of parametrized Minkowski theories it has been possible to
reformulate relativistic fluids \cite{45,46,47} (starting from
the viewpoint of Ref.\cite{48}) and classical fields in non-inertial frames
and in the rest-frame instant form with the only condition that
the 10 conserved generators of the Poincar\'e algebra are finite. In Ref. \cite{49} this is done
for the Klein-Gordon field and for it a canonical basis containing the
relativistic collective center-of-mass and relative variables was found.
The same was done for the Dirac equation in Refs.\cite{50,51} after the elimination of the existing
second class constraints implied by the first order Dirac Lagrangian.

\medskip

Inspired by Dirac \cite{52}, who showed that the DO's of
the electromagnetic field are the transverse vector potential
${\vec A}_{\perp}$ and the transverse electric field ${\vec
E}_{\perp}$ of the {\it radiation gauge}, we describe Maxwell theory
in the rest frame in Subsection A. This formulation has been used
in the semi-classical approximation to get a consistent
description of N charged particles plus the electromagnetic field \cite{19,24},
if we use Grassmann -valued electric charges to regularize the
Coulomb self-energies. In Ref.\cite{25,26,27} the electromagnetic
degrees of freedom are expressed in terms of the particle
variables by means the Lienard-Wiechert solution and this allows
to find the relativistic Darwin potential (or the Salpeter
potential for spinning particles) starting from classical
electrodynamics and not as a reduction from QFT.
In these papers it is shown that when a fermion field is interacting with the
electromagnetic field, the fermionic DO is
{\it a fermion field dressed with a Coulomb cloud}.

\bigskip

Then in Subsection B we will look at the Yang-Mills field whose rest-frame
formulation was done in Ref.\cite{53} in the framework of the quark model (with scalar quarks).
This paper puts in Wigner-covariant form the results of previous papers on the use of
constraint theory for the description of Yang-Mills theory with fermions \cite{54},
of Higgs models \cite{55} and of the $SU(3) \times SU(2) \times U(1)$ model \cite{56}.

\subsection{The Electro-Magnetic Field and its Dirac Observables}

As shown in Refs.\cite{19,24} in parametrized Minkowski theories
the configuration variable for the electro-magnetic field is  the Lorentz-scalar
potential $A_A(\tau ,\vec \sigma ) = z^{\mu}_A(\tau, \vec \sigma)\, {\tilde
A}_{\mu}(z^{\beta}(\tau ,\vec \sigma ))$, whose associated field
strength is $F_{AB}(\tau ,\vec \sigma ) = \partial_A\, A_B(\tau ,\vec \sigma ) -
\partial_B\, A_A(\tau ,\vec \sigma ) = z^{\mu}_A(\tau ,\vec \sigma
)\, z^{\nu}_B(\tau ,\vec \sigma )\, {\tilde F}_{\mu\nu}(z^{\beta}(\tau ,\vec
\sigma ))$. After the restriction to the inertial rest frame
the conjugate electro-magnetic momentum variables are a scalar
$\pi^{\tau}(\tau ,\vec \sigma )$ and a Wigner 3-vector
$\pi^r(\tau ,\vec \sigma ) = E^r(\tau ,\vec \sigma )$.
The primary first-class constraint is $\pi^{\tau}(\tau ,\vec \sigma ) \approx 0$,
while the secondary first-class constraint is
the Gauss law $\Gamma(\tau ,\vec \sigma )
= \vec \partial \cdot {\vec \pi}(\tau ,\vec \sigma ) \approx 0$.
They are the generators of the Hamiltonian electro-magnetic gauge transformations.
$E^r(\tau ,\vec \sigma )$ and $B^r(\tau ,\vec \sigma ) = (\vec \partial
\times {\vec A}_{\perp})^r(\tau, \vec \sigma)$ are the components of
the electric and magnetic fields.
\medskip

The gauge degrees of freedom ($A_{\tau}$, $\eta$) have been
separated from the transverse DO's of Ref.\cite{52} ($A_{\perp r}$, $
\pi^r_{\perp} = E^r_{\perp}$) ($\vec \partial \cdot {\vec A}_{\perp} =
\vec \partial \cdot {\vec \pi}_{\perp} = 0$) by means of a Shanmughadhasan
canonical transformation \cite{19,24,57}  adapted to the two scalar first
class constraints  ($\triangle = -
{\vec \partial}^2_{\sigma}$, $\Box = \partial^2_{\tau} + \triangle$)

\begin{eqnarray*}
 &&\begin{minipage}[t]{1cm}
\begin{tabular}{|l|} \hline
$A_A$ \\  \hline
 $\pi^A$ \\ \hline
\end{tabular}
\end{minipage} \ {\longrightarrow \hspace{.2cm}} \
\begin{minipage}[t]{2 cm}
\begin{tabular}{|l|l|l|} \hline
$A_{\tau}$ & $\eta$   & $A_{\perp\, r}$   \\ \hline
$\pi^{\tau}\approx 0$& $\Gamma \approx 0$ &$\pi^r_{\perp}$ \\
\hline
\end{tabular}
\end{minipage}
 \end{eqnarray*}

 \bea
 A_r(\tau ,\vec \sigma )&=&
\partial_r\, \eta (\tau ,\vec \sigma
)+A^r_{\perp}(\tau ,\vec \sigma ),\qquad
 \pi^r(\tau ,\vec \sigma ) = \pi^r_{\perp}(\tau
,\vec \sigma )+{1\over {\triangle_{\sigma}} }{ {\partial}\over
{\partial \sigma^r}} \, \Gamma (\tau ,\vec \sigma ), \nonumber \\
 &&{}\nonumber \\
  \eta(\tau ,\vec \sigma )&=&-{1\over {\triangle_{\sigma}} }{ {\partial} \over
 {\partial \vec \sigma} }\cdot \vec A(\tau ,\vec \sigma ),\nonumber \\
 &&{}\nonumber \\
 A^r_{\perp}(\tau ,\vec \sigma ) &=& (\delta^{rs}+{{\partial^r_{\sigma}\partial^s_{\sigma}}
 \over {\triangle_{\sigma} }})\, A_s(\tau ,\vec \sigma ),\qquad
  \pi^r_{\perp}(\tau ,\vec \sigma ) = (\delta^{rs}+{{\partial^r_{\sigma}\partial^s_{\sigma}}
 \over {\triangle_{\sigma} }})\, \pi_s(\tau ,\vec \sigma ),\nonumber \\
 &&{}\nonumber \\
  &&\lbrace A_{\tau}(\tau ,\vec \sigma ), \pi^{\tau}
(\tau ,{\vec \sigma}^{'} ) \rbrace = -
  \lbrace \eta (\tau ,\vec \sigma ),\Gamma
(\tau ,{\vec \sigma}^{'} ) \rbrace =  \delta^3(\vec
\sigma -{\vec \sigma}^{'}),\nonumber \\
 &&{}\nonumber \\
 &&\lbrace A^r_{\perp}(\tau ,\vec
\sigma ),\pi^s_{\perp}(\tau ,{\vec \sigma} ^{'})\rbrace =-
(\delta^{rs}+{{\partial^r_{\sigma}\partial^s_{\sigma}} \over
{\triangle_{\sigma} }})\delta^3(\vec \sigma -{\vec \sigma}^{'}).
 \label{3.1}
 \eea

The Dirac Hamiltonian is ($\lambda_{\tau}(\tau ,\vec \sigma )$ is
the arbitrary Dirac multiplier associated to the primary constraint
$\pi^{\tau}(\tau ,\vec \sigma ) \approx 0$)

 \bea
 H_D &=& H_c + \int d^3\sigma\, [\lambda_{\tau}\, \pi^{\tau}-
 A_{\tau}\, \Gamma ](\tau ,\vec \sigma ),\qquad
 H_c = {1\over 2}\, \int d^3\sigma\, [{\vec \pi}^2_{\perp}
 + {\vec B}^2](\tau ,\vec \sigma ),\nonumber \\
 &&{}\nonumber \\
 \Downarrow &&kinematical\, Hamilton \, equations\nonumber \\
 &&{}\nonumber \\
 \partial_{\tau}\, A_{\tau}(\tau ,\vec \sigma )\, &\cir&
 \lambda_{\tau}(\tau ,\vec \sigma ),\qquad
 \partial_{\tau}\, \eta (\tau ,\vec \sigma )\, \cir\, A_{\tau}(\tau ,\vec \sigma
 ),\qquad \partial_{\tau}\, A_{\perp\, r}(\tau ,\vec \sigma )\, \cir\,
 - \pi_{\perp\, r}(\tau ,\vec \sigma ),\nonumber \\
 &&{}\nonumber \\
 &&dynamical\, Hamilton\, equations\nonumber \\
 &&{}\nonumber \\
 \partial_{\tau}\, \pi^r_{\perp}(\tau ,\vec \sigma )\, &\cir&\,
 \triangle\, A^r_{\perp}(\tau ,\vec \sigma ),\quad
 \Rightarrow\quad \Box\, A_{\perp\, r}(\tau ,\vec \sigma )\,
 \cir\, 0.
 \label{3.2}
 \eea
\bigskip

To fix the gauge we must only add a gauge fixing
$\varphi_{\eta}(\tau ,\vec \sigma ) \approx 0$ to the Gauss law,
which determines $\eta$. Its time constancy, i.e. $\partial_{\tau}\,
\varphi_{\eta}(\tau ,\vec \sigma ) + \{ \varphi_{\eta}(\tau ,\vec
\sigma ), H_D\} =  \varphi_{A_{\tau}}(\tau
,\vec \sigma ) \approx 0$, will generate the gauge fixing
$\varphi_{A_{\tau}}(\tau ,\vec \sigma ) \approx 0$ for $A_{\tau}$
(as shown in Ref.\cite{58} this is the correct procedure for fixing a gauge).
Finally the time constancy $\partial_{\tau}\,
\varphi_{A_{\tau}}(\tau ,\vec \sigma ) + \{ \varphi_{A_{\tau}}(\tau
,\vec \sigma ), H_D\} \approx 0$ will determine the Dirac multiplier
$\lambda_{\tau}(\tau ,\vec \sigma )$. By adding these {\it
two}  gauge fixing constraints  to the
first class constraints $\pi^{\tau}(\tau ,\vec \sigma ) \approx 0$,
$\Gamma (\tau ,\vec \sigma ) \approx 0$, one gets two pairs of
second class constraints allowing the elimination of the gauge
degrees of freedom so that only the DO's survive.
\bigskip

As shown in Refs.\cite{19,24} one can add  point particles to define
relativistic atomic physics and to study relativistic bound states.
Moreover as shown in Ref.\cite{59} the standard Lagrangian density
for the electromagnetic field coupled to Dirac fields
 ${\cal L}(x)=-{1\over 4}\, F^{\mu\nu}(x)\,
F_{\mu\nu}(x)+\bar \psi (x)\, \gamma^{\mu}\, (i\partial_{\mu}+e\,
A_{\mu}(x))\, \psi (x)-m\, \bar \psi (x)\, \psi (x)-
\partial_{\mu}\, ({i\over 2}\, \bar \psi (x)\, \gamma^{\mu}\,
\psi (x))$ has a generalized weak quasi-invariance: $\delta\, {\cal
L} = (\eta - \partial_o\, \epsilon )\, \Gamma \cir 0$ under
the generalized Hamiltonian gauge transformations
$\delta\, A_o = \eta$ and $\delta\, A_k = \partial_k\, \epsilon$
generated by the electro-magnetic first class constraints.
As a consequence the Noether identities implied by the second Noether theorem
reproduce Dirac's algorithm and one can show that
the {\it improper strong conserved electric charge} $Q^{(S)}$
(the flux through the surface at infinity of the electric field) and the
{\it improper weak conserved Noether electric charge} $Q^{(W)}$ (the
volume integral over the charged fermion density) coincide due to the first class
secondary Gauss law constraint:
$Q^{(S)} = \int_{\partial \Omega}\, d^{m-1}\Sigma_k\, E^k(x^o, \vec x)
= \int_{\Omega}\, d^mx\, \Big[ e\, [\psi^{\dagger}\, \psi ] (x^o, \vec x) + \Gamma(x^o, \vec
x)\Big]  \cir Q^{(W)} =  \int_{\Omega}\, d^mx\, e\, [\psi^{\dagger}\, \psi ](x^o, \vec x)$.

\subsection{Yang-Mills Fields }

Notwithstanding Eqs.(\ref{3.1}) for the electro-magnetic case,
in general gauge field theories the situation is more complicated, because
some of the constraints are non-linear partial differential equations (PDE) and
the theorems ensuring the existence of the Shanmugadhasan
canonical transformation have not been extended to the
infinite-dimensional case so that one must use heuristic
extrapolations of them. In  Yang-Mills theory with gauge potentials
$A_\mu (x)=A^a_\mu (x)T_a$ \footnote{$T_a$ are the generators of the Lie algebra
$g$ of the Lie group G with
$T_a^\dagger =-T_a$, $Tr(T_aT_b)=-\delta_{ab}$,
$[T_a,T_b]=c_{abc}T_c$. The field strengths are  $F_{\mu\nu}=F^a_{\mu\nu}T_a=\partial_\mu
A_\nu -\partial_\nu A_\mu+[A_\mu,A_\nu]$ and the Lagrangian is ${\cal L}={1\over
2}Tr(F^{\mu\nu}F_{\mu\nu})$. The canonical momenta are
$\pi^{\mu a}=F^{\mu oa}$. The EL equations are
$L^{\mu} = L^{\mu\, a}\, T_a = D_{\nu}\, F^{\nu\mu} =
\partial_{\nu}\, F^{\nu\mu} + [ A_{\nu}, F^{\nu\mu} ] \cir 0$.
The primary first-class constraints are $\pi^o_a(x^o, \vec x) \approx 0$,
while the secondary ones are the Gauss laws $\Gamma_a(x^o, \vec x) = L^éoa* \approx 0$.
The infinitesimal gauge transformations under which $\delta {\cal
L}\equiv 0$ are $\delta A_{\mu a}(x)
=\partial_\mu\epsilon_a(x)+c_{abc}A_{\mu b}\, \epsilon_c(x)$.}
some of the constraints  are elliptic PDE's and they
can have {\it zero modes}. Let us consider the stratum $\sgn\, p^2 > 0$
of free Yang-Mills theory as a prototype and its first class
constraints, given by the Gauss laws and by the vanishing of the
time components of the canonical momenta. The problem of the {\it
zero modes} will appear as a {\it singularity structure of the
gauge foliation} of the allowed strata, in particular of the
stratum $\sgn\, p^2 > 0$. This phenomenon was discovered in
Refs.\cite{60,61} by studying the space of solutions of Yang-Mills
and Einstein equations, which can be mapped onto the constraint
manifold of these theories in their Hamiltonian description. It
turns out that the space of solutions has a {\it cone over cone}
structure of singularities: if we have a line of solutions with a
certain number of symmetries, in each point of this line there is
a cone of solutions with one less symmetry.

\medskip

Other possible sources of singularities of the gauge foliation of
Yang-Mills theory in the stratum $\sgn\, p^2 > 0$ may be:
\noindent i) different classes of gauge potentials identified by
different values of the field invariants;
\noindent ii) the orbit structure of the rest frame (or Thomas)
spin $\vec S$, identified by the Pauli-Lubanski Casimir $W^2 = -
\sgn\, p^2\, {\vec S}^2$ of the Poincare' group.
The final outcome of this structure of singularities is that the
reduced phase-space, i.e. the space of the gauge orbits, is in
general a {\it stratified manifold with singularities} \cite{60,61}.
In the stratum $\sgn\, p^2 > 0$ of the Yang-Mills theory these
singularities survive the Wick rotation to the Euclidean
formulation and it is not clear how the ordinary path integral
approach and the associated BRS method can take them into account
(they are zero measure effects).

\medskip

In the Yang-Mills case there is also the problem of the {\it  gauge symmetries}
of a gauge potential $A_\mu (x)=A^a_\mu (x)T_a$, which
are connected with the generators of its stability group, i.e. with the subgroup
of those special gauge transformations which leave invariant that
gauge potential. This is the {\it Gribov ambiguity} for gauge
potentials \cite{62} (see Ref.\cite{63} for a recent contribution to its
mathematical aspects). There is also a more general Gribov ambiguity for
field strengths, the {\it gauge copies} problem due to those gauge transformations
leaving invariant the field strengths.
For all these problems see Ref. \cite{54} and its bibliography.

Since the Gauss laws $\Gamma_a(x^o, \vec x) = D^{(A)}_{ab}(x^0, \vec x) \cdot
{\vec \pi}_b(x^o, \vec x) = \partial_r \pi^r_a(x^o, \vec x) + c_{abc}\, A_{br}(x^o,
\vec x)\, \pi^r_c(x^o, \vec x) \approx 0$
($\{ \Gamma_a(x^o, \vec x), \Gamma_b(x^o, \vec y) \} = c_{abc}\, \Gamma_c(x^o,
\vec x)\, \delta^3(\vec x - \vec y)$)
are the generators of the gauge
transformations (and depend on the chosen gauge potential through
the covariant derivative), this means that for a gauge potential
with non trivial stability group those combinations of the Gauss
laws corresponding to the generators of the stability group {\it cannot
be any more first class constraints}, since they do not generate
effective gauge transformations but special symmetry
transformations. This problematic has still to be clarified, but
it seems that in this case these components of the Gauss laws
become {\it third class constraints}, which are not generators of
true gauge transformations. This new kind of constraints was
introduced in Refs.\cite{57,59} in the finite dimensional case as
a result of the study of some examples, in which the Jacobi
equations (the linearization of the Euler-Lagrange equations) are
singular, i.e. some of their solutions are not infinitesimal
deviations between two neighboring extremals of the Euler-Lagrange
equations. This interpretation seems to be confirmed by the fact
that the singularity structure discovered in Ref.\cite{60,61}
follows from the existence of singularities of the linearized
Yang-Mills and Einstein equations.
Due to the Gribov ambiguity, to fix univocally a gauge one has to give topological
numbers identifying a stratum besides the ordinary gauge fixing constraints.

\medskip

The search of a global canonical
basis of DO's for each stratum of the space of the gauge orbits can
give a definition of the measure of the phase space path integral,
but at the price of a non polynomial Hamiltonian. Therefore, if it
is not possible to eliminate the Gribov ambiguity (assuming that
it is only a mathematical obstruction without any hidden physics
like in the approach to QCD confinement reviewed in Ref.\cite{64}),
the existence of global DO's for Yang-Mills theory is very
problematic.

\bigskip

In Ref.\cite{54} there is the study of Yang-Mills theory with Grassmann-valued fermion fields
in the case of a trivial principal bundle over a
fixed-$x^o$ $R^3$ slice of Minkowski space-time with suitable
Hamiltonian-oriented boundary conditions; this excludes monopole
solutions and, since $R^3$ is not compactified, one has only
winding number and no instanton number. After a discussion of the
Hamiltonian formulation of Yang-Mills theory
(with a suitable coordinatization of the Lie group and of the principal bundle), of its group of
gauge transformations and of the Gribov ambiguity, the theory has been studied in suitable
{\it weighted Sobolev spaces} where the Gribov ambiguity is absent \cite{65,66}
and the global color charges are well defined.
In this case it is possible to find the non-Abelian analogue of the Abelian
Shanmugadhasan canonical transformation (\ref{3.1}) in an inertial
frame of Minkowski space-time (these results have been rewritten in the rest-frame
instant form like in the electro-magnetic case in Ref.\cite{53}), namely gauge
variables $A^o_a(x^o, \vec x)$ and $\eta_a(x^o, \vec x)$ canonically conjugated to the first-class
constraints $\pi_{oa}(x^o, \vec x) \approx 0$ and to  Abelianized Gauss law
constraints ${\tilde \Gamma}_a(x^o, \vec x) \approx 0$
($\{ {\tilde \Gamma}_a(x^o, \vec x), {\tilde \Gamma}_b(x^o, \vec y) \} = 0$).
Moreover there are the global DO's, i.e.
transverse quantities ${\vec A}_{a\perp} (\vec x,x^o)$, ${\vec
E}_{a\perp}(\vec x,x^o)$ (and fermion fields dressed with
Yang-Mills (gluonic) clouds). All these quantities are extremely complex,
non-local and with a poor global control due to the need of the
Green function of the covariant divergence (it requires the use of path-dependent
non-integrable phases; see Ref.\cite{53}). The nonlocal and non-polynomial (due
to the presence of classical Wilson lines along flat geodesics)
physical Hamiltonian has been obtained: it is nonlocal but without
any kind of singularities and it has the correct Abelian limit if the
structure constants are turned off.

\bigskip

Again the Noether identities implied by the second Noether theorem \cite{59}
reproduce the Dirac algorithm for the identification of the constraints.
In the suitable weighted Sobolev spaces eliminating the Gribov ambiguity
the strong conserved charge is $V^{\mu a}=\partial_\nu F^{\nu\mu
a}=\partial_\nu U^{[\mu\nu]a}$, with the super-potential
$U^{[\mu\nu]a}=-F^{\mu\nu a}$ and the improper strong and weak
conservation laws are $\partial_\mu V^{\mu a}\equiv 0$ and
$\partial_\mu G^\mu_{1a}\cir 0$ respectively. The improper strong
and weak conserved non-Abelian color charges (in absence of fermions;
they would add a term like the one in the Abelian electric charge) are
$Q^{(S)}_a  = \int_{\partial \Omega}\,  d^{m-1}\Sigma_k\, F^{ko}_a \cir
Q^{(W)}_a= -C_{abc}\, \int_\Omega\, d^mx\, F^{ok}_b\, A_{k c}$.

\bigskip

The following models have been studied with the described technology:

A) SU(3) Yang-Mills theory with scalar particles with
Grassmann-valued color charges \cite{53} for the regularization
of self-energies. It is possible to show that in this relativistic
scalar quark model the Dirac Hamiltonian expressed as a function
of DO's has the property of asymptotic freedom.

B) The Abelian and non-Abelian SU(2) Higgs models with fermion
fields \cite{55}, where the symplectic decoupling is a refinement
of the concept of unitary gauge. There is an ambiguity in the
solutions of the Gauss law constraints, which reflects the
existence of disjoint sectors of solutions of the Euler-Lagrange
equations of Higgs models. The physical Hamiltonian and Lagrangian
of  the Higgs phase have been found; the self-energy turns out to
be local and contains a local four-fermion interaction.

C) The standard SU(3)xSU(2)xU(1) model of elementary particles
\cite{56} with Grassmann- valued fermion fields. The
final reduced Hamiltonian contains nonlocal self-energies for the
electromagnetic and color interactions, but "local ones" for the
weak interactions implying the non-perturbative emergence of
4-fermions interactions.

\section{Einstein General Relativity, Tetrad Gravity and their Canonical ADM Formulation}

In this Section we introduce the Hamiltonian description of Einstein's GR.
Instead of studying ADM metric gravity \cite{67} we will look at tetrad gravity \cite{68}, because it is  needed for
the descriptions of the fermions of the particle standard model.

The use of Hamiltonian methods restricts the class of Einstein
space-times to the {\it globally hyperbolic} ones, in which there is
a global notion of a mathematical time parameter and of instantaneous
3-spaces to be used as Cauchy surfaces for the dynamics. The space-times
must also be {\it topologically trivial}. Moreover the space-times
must be asymptotically Minkowskian space-times without super-translations so to have the
asymptotic ADM Poincare' group  \cite{67,69} (needed for particle physics, where elementary particles are always defined as
irreducible representations of the Poincare' group and all the properties are  connected with the
representations of this group in the inertial frames of Minkowski space-time), absent in
loop quantum gravity where the 3-spaces are compact manifolds without boundary \cite{9,10}.

In Subsection A we will describe the global non-inertial frames of the space-times. In Subsection B
there is the parametrization of ADM tetrad gravity and in Subsection C its Hamiltonian formulation.
In Subsection D there is a partial Shanmugadhasan canonical transformation adapted to 10 of the 14 first class
constraints with a clarification on which are the inertial gauge variables and the tidal
physical degrees of freedom of GR. In Subsection E there are comments on the DO's of GR
and on the attempts of its quantization. In Subsection G there is the description of the
3-orthogonal Schwinger time gauges, replacing the harmonic ones in this framework to describe
post-Minkowskian gravity in presence of point particles and electro-magnetic fields and its
possible relevance  for the problem of dark matter.

\subsection{3+1 Splittings of Asymptotically Minkowskian Space-Times}

In asymptotically flat space-times we have the
asymptotic symmetries of the SPI group \cite{70}
(direction-dependent asymptotic Killing symmetries). If we restrict
this class of space-times to those {\it not containing
super-translations} \cite{69}, the SPI group reduces to the {\it
asymptotic ADM Poincar\'e group} \footnote{For recent reviews on
this group see Refs.\cite{71,72,73}.}: these space-times are
asymptotically Minkowskian \footnote{This class of space-times admits
ortho-normal tetrads and a spinor structure \cite{74}.}  and in the limit of vanishing Newton
constant ($G = 0$) the ADM Poincar\'e group becomes the special
relativistic Poincar\'e group of the matter present in the
space-time. In this restricted class the
canonical Hamiltonian is the ADM energy \cite{75}, so that there is
no frozen picture (in the reduced phase space there is a non-zero
reduced Hamiltonian). In absence of matter a sub-class of these
space-times is the (singularity-free) family of
Chrstodoulou-Klainermann solutions of Einstein equations \cite{76}
(they are near to Minkowski space-time in a norm sense and contain
gravitational waves).

\medskip

In the considered class of Einstein space-times the ten {\it strong} asymptotic ADM
Poincar\'e generators $P^A_{ADM}$, $J^{AB}_{ADM}$ (they are fluxes
through a 2-surface at spatial infinity) are well defined
functionals of the 4-metric fixed by the boundary conditions at
spatial infinity and of matter (when present). These ten strong generators can be
expressed \cite{69,75}  in terms of the weak asymptotic ADM Poincar\'e generators
${\hat P}^A_{ADM}$, ${\hat J}^{AB}_{ADM}$  (integrals on
the 3-space of suitable densities) plus first class constraints.
The absence of super-translations implies that the ADM 4-momentum is asymptotically orthogonal to the
instantaneous 3-spaces (they tend to a Euclidean 3-space at spatial infinity), namely
there is an asymptotic rest frame condition ${\hat P}^r_{ADM} \approx 0$. As a consequence each
3-space of the global non-inertial frame is a {\it non-inertial rest frame} of the 3-universe. At spatial
infinity there are asymptotic inertial observers carrying a flat tetrad whose spatial axes are identified by the fixed stars of star catalogues.

\medskip

Instead in spatially compact space-times without boundary
the canonical Hamiltonian is zero and the Dirac Hamiltonian is a
linear combination of first class constraints. This fact gives rise
to a {\it frozen picture} without a global evolution (the Dirac
Hamiltonian generates only Hamiltonian gauge transformations; in the
abstract reduced phase space, quotient with respect to such gauge
transformations, the reduced Hamiltonian is zero). This class of
space-times fits well with Machian ideas (no boundary conditions),
with interpretations in which there is no physical time (see for
instance Ref.\cite{77}) and is used in loop quantum gravity \cite{9,10}.

\medskip

If a space-time  without non-asymptotic Killing symmetries and with
the fields belonging to suitable weighted Sobolev spaces is globally hyperbolic,
topologically trivial, asymptotically Minkowskian and without super-translations
then there is a well established Hamiltonian description
of both metric and tetrad gravity \cite{75,78,79,80,81,82,83}
(see Refs.\cite{31,84}  for  reviews). This is due to the fact that
in these space-times one can make a consistent 3+1 splitting with instantaneous
non-Euclidean 3-spaces (i.e. a clock synchronization convention \cite{16}) centered
on a time-like observer used as origin of (world scalar)
radar 4-coordinates \cite{17,18}: in this way the notion of non-inertial frames
defined in Minkowski space-time described in the previous Section
can be extended to this class of curved space-times, where the equivalence principle forbids the existence of global
inertial frames.
\medskip

In GR the dynamical fields are the components ${}^4g_{\mu\nu}(x)$ of
the 4-metric and not the  embeddings $x^{\mu} = z^{\mu}(\tau,
\vec \sigma)$ defining the admissible 3+1 splittings of space-time. Now the gradients
$z^{\mu}_A(\tau, \vec \sigma)$ of the embeddings give the transition
coefficients from radar to world 4-coordinates, so that the
components ${}^4g_{AB}(\tau, \vec \sigma) = z^{\mu}_A(\tau, \vec \sigma)\,
z^{\nu}_B(\tau, \vec \sigma)\, {}^4g_{\mu\nu}(z(\tau, \vec \sigma))$ of
the 4-metric will be the dynamical fields in the ADM action \cite{67}.
Let us remark that {\it the ten quantities ${}^4g_{AB}(\tau, \vec \sigma)$ are
4-scalars of the space-time due to the use of the world-scalar radar 4-coordinates}. In each
3-space $\Sigma_{\tau}$ considered as a 3-manifold with 3-coordinates $\sigma^r$ (and
not as a 3-sub-manifold of the space-time) ${}^4g_{\tau r}(\tau, \vec \sigma)$ is a 3-vector
and ${}^4g_{rs}(\tau, \vec \sigma)$ is a 3-tensor. Therefore {\it all the components of  "radar tensors",
i.e. tensors expressed in radar 4-coordinates, are 4-scalars of the space-time} \cite{83}.

\bigskip

Flat indices $(\alpha )$, $\alpha = o, a$,
are raised and lowered by the flat Minkowski metric
${}^4\eta_{(\alpha )(\beta )} = \sgn\, (+---)$. We define
${}^4\eta_{(a)(b)} = - \sgn\, \delta_{(a)(b)}$ with a
positive-definite Euclidean 3-metric.

\subsection{ADM Tetrad Gravity}

In tetrad gravity   the 4-metric is decomposed in terms of cotetrads,
${}^4g_{AB} = E_A^{(\alpha)}\, {}^4\eta_{(\alpha)(\beta)}\,
E^{(\beta)}_B$ and the ADM action, now a
functional of the 16 fields $E^{(\alpha)}_A(\tau, \vec \sigma)$, is
taken as the action for ADM tetrad gravity. The
diffeomorphism group (the gauge group of GR) is enlarged with the O(3,1) gauge group of the
Newman-Penrose approach \cite{85} (the extra gauge freedom acting
on the tetrads in the tangent space of each point of space-time and
reducing from 16 to 10 the number of independent fields  in the restriction to
metric gravity). This leads to an interpretation of gravity based on
a congruence of time-like observers endowed with ortho-normal
tetrads: in each point of space-time the time-like axis is the  unit
4-velocity of the observer, while the spatial axes are a (gauge)
convention for observer's gyroscopes. This framework was developed
in Refs.\cite{75,78}.

\medskip

In this framework the configuration variables are {\it cotetrads}, which
are connected to cotetrads adapted to the 3+1 splitting of
space-time (so that the adapted time-like tetrad is the unit normal
$l^{\mu}(\tau, \vec \sigma) = \Big(z^{\mu}_A\, l^A\Big)(\tau, \vec
\sigma)$ to the 3-space $\Sigma_{\tau}$) by standard Wigner boosts for
time-like vectors \footnote{In each tangent plane to a point of
$\Sigma_{\tau}$ the point-dependent standard Wigner boost for
time-like Poincare' orbits $L^{(\alpha )}{}_{(\beta )}(V(z(\sigma
));\,\, {\buildrel \circ \over V}) = \delta^{(\alpha )}_{(\beta )} +
2 \sgn\, V^{(\alpha )}(z(\sigma ))\, {\buildrel \circ \over
V}_{(\beta )} - \sgn\, {{(V^{(\alpha )}(z(\sigma )) + {\buildrel
\circ \over V}^{(\alpha )})\, (V_{(\beta )}(z(\sigma )) + {\buildrel
\circ \over V}_{(\beta )})}\over {1 + V^{(o)}(z(\sigma ))}}\,
{\buildrel {def}\over =}\, L^{(\alpha )}{}_{(\beta
)}(\varphi_{(a)})$ sends the unit future-pointing time-like vector
${\buildrel o\over V}^{(\alpha )} = (1; 0)$ into the unit time-like
vector $V^{(\alpha )} = {}^4E^{(\alpha )}_A\, l^A = \Big(\sqrt{1 +
\sum_a\, \varphi^2_{(a)}}; \varphi^{(a)} = - \sgn\,
\varphi_{(a)}\Big)$, where $l^A$ is the unit future-pointing normal
to $\Sigma_{\tau}$. We have $L^{-1}(\varphi_{(a)}) = {}^4\eta\,
L^T(\varphi_{(a)})\, {}^4\eta = L(- \varphi_{(a)})$. As a
consequence, the flat indices $(a)$ of the adapted tetrads and
cotetrads and of the triads and cotriads on $\Sigma_{\tau}$
transform as Wigner spin-1 indices under point-dependent SO(3)
Wigner rotations $R_{(a)(b)}(V(z(\sigma ));\,\, \Lambda (z(\sigma
))\, )$ associated with Lorentz transformations $\Lambda^{(\alpha
)}{}_{(\beta )}(z)$ in the tangent plane to the space-time in the
given point of $\Sigma_{\tau}$. Instead the index $(o)$ of the
adapted tetrads and cotetrads is a local Lorentz scalar index.} of
parameters $\varphi_{(a)}(\tau, \vec \sigma)$:
${}^4E_A^{(\alpha)} = L^{(\alpha)}{}_{(\beta)}( \varphi_{(a)})\,
{}^4{\buildrel o\over E}_A^{(\beta)}$.
The {\it adapted tetrads and cotetrads}   have the expression
$\eo^A_{(o)} = {1\over {1 + n}}\, (1; - \sum_a\, n_{(a)}\,
{}^3e^r_{(a)}) = l^A$, $\eo^A_{(a)} = (0; {}^3e^r_{(a)})$,
$\eo^{(o)}_A = (1 + n)\, (1; \vec 0) = \sgn\, l_A$, $\eo^{(a)}_A
= (n_{(a)}; {}^3e_{(a)r})$, where ${}^3e^r_{(a)}$ and ${}^3e_{(a)r}$ are {\it triads and
cotriads} on $\Sigma_{\tau}$  (since we use the positive-definite
3-metric $\delta_{(a)(b)} $, we shall use only lower flat spatial
indices: therefore for the cotriads we use the notation
${}^3e^{(a)}_r\,\, {\buildrel {def}\over =}\, {}^3e_{(a)r}$ with
$\delta_{(a)(b)} = {}^3e^r_{(a)}\, {}^3e_{(b)r}$). The lapse and shift functions are
$N(\tau, \vec \sigma) = 1 + n(\tau, \vec \sigma) > 0$ and
$N^r(\tau, \vec \sigma) = n^r(\tau, \vec \sigma)$ with
$n(\tau ,\vec \sigma)$  and $n^r(\tau, \vec \sigma)$
vanishing at spatial infinity due to the absence of super-translations \cite{78};
$n_{(a)} = n_r\, {}^3e^r_{(a)} = n^r\, {}^3e_{(a)r}$ are adapted shift
functions.
The adapted tetrads $\eo^A_{(a)}$ are defined modulo SO(3) rotations
$\eo^A_{(a)} = \sum_b\, R_{(a)(b)}(\alpha_{(e)})\, {}^4{\buildrel
\circ \over {\bar E}}^A_{(b)}$, ${}^3e^r_{(a)} = \sum_b\,
R_{(a)(b)}(\alpha_{(e)})\, {}^3{\bar e}^r_{(b)}$, where
$\alpha_{(a)}(\tau ,\vec \sigma )$ are three point-dependent Euler
angles. After having chosen an arbitrary point-dependent origin
$\alpha_{(a)}(\tau ,\vec \sigma ) = 0$, we arrive at the following
{\it adapted tetrads and cotetrads} [${\bar n}_{(a)} = \sum_b\, n_{(b)}\,
R_{(b)(a)}(\alpha_{(e)})\,$, $\sum_a\, n_{(a)}\, {}^3e^r_{(a)} =
\sum_a\, {\bar n}_{(a)}\,
 {}^3{\bar e}^r_{(a)}$]

\bea
 {}^4{\buildrel \circ \over {\bar E}}^A_{(o)}
 &=& \eo^A_{(o)} = {1\over {1 + n}}\, (1; - \sum_a\, {\bar n}_{(a)}\,
 {}^3{\bar e}^r_{(a)}) = l^A,\qquad {}^4{\buildrel \circ \over
 {\bar E}}^A_{(a)} = (0; {}^3{\bar e}^r_{(a)}), \nonumber \\
 &&{}\nonumber  \\
 {}^4{\buildrel \circ \over {\bar E}}^{(o)}_A
 &=& \eo^{(o)}_A = (1 + n)\, (1; \vec 0) = \sgn\, l_A,\qquad
 {}^4{\buildrel \circ \over {\bar E}}^{(a)}_A
= ({\bar n}_{(a)}; {}^3{\bar e}_{(a)r}),
 \label{4.1}
 \eea

\noindent which we shall use as a {\it reference standard}.
We have ${}^4g_{AB} = \eo^{(\alpha)}_A\, {}^4\eta_{(\alpha )(\beta )}\, \eo^{(\beta )}_B =
{}^4{\buildrel \circ \over {\bar E}}^{(\alpha)}_A\,
{}^4\eta_{(\alpha)(\beta)}\, {}^4{\buildrel \circ \over {\bar E}}^{(\beta)}_B$.
The expressions for the general tetrad and for the 4-metric
(the 3-metric ${}^3g_{rs}$ has signature $(+++)$, so that we may put
all the flat 3-indices {\it down}; we have ${}^3g^{ru}\, {}^3g_{us}
= \delta^r_s$) are

\bea
 {}^4E^A_{(\alpha )} &=& \eo^A_{(\beta )}\, L^{(\beta )}{}_{(\alpha
 )}(\varphi_{(a)}) = {}^4{\buildrel \circ \over {\bar E}}^A_{(o)}\,
 L^{(o)}{}_{(\alpha )}(\varphi_{(c)}) + \sum_{ab}\, {}^4{\buildrel \circ \over
 {\bar E}}^A_{(b)}\, R^T_{(b)(a)}(\alpha_{(c)})\,
 L^{(a)}{}_{(\alpha )}(\varphi_{(c)}),\nonumber \\
   {}^4g_{\tau\tau} &=& \sgn\, [(1 + n)^2 - {}^3g^{rs}\, n_r\,
 n_s] = \sgn\, [(1 + n)^2 - \sum_a\, {\bar n}^2_{(a)}],\qquad
 {}^4g_{\tau r} = - \sgn\, n_r = -\sgn\, \sum_a\, {\bar n}_{(a)}\,
 {}^3{\bar e}_{(a)r},\nonumber \\
  {}^4g_{rs} &=& -\sgn\, {}^3g_{rs} = - \sgn\, \sum_a\, {}^3e_{(a)r}\, {}^3e_{(a)s}
  = - \sgn\, \sum_a\, {}^3{\bar e}_{(a)r}\, {}^3{\bar e}_{(a)s},\nonumber \\
   \sqrt{- g } &=& \sqrt{|{}^4g|} = {{\sqrt{{}^3g}}\over {\sqrt{\sgn\,
 {}^4g^{\tau\tau}}}} = \sqrt{\gamma}\, (1 + n) = {}^3e\, (1 +
 n),\qquad {}^3g = \gamma =
 ({}^3e)^2,\quad {}^3e = det\, {}^3e_{(a)r}.\nonumber \\
 &&{}
  \label{4.2}
 \eea

\bigskip

Each 3+1 splitting of an either Minkowski or asymptotically
Minkowskian space-time, i.e. each global non-inertial frame, has
two associated {\it congruences of time-like observers}:
i) The congruence of the Eulerian observers with the unit normal
$l^{\mu}(\tau, \vec \sigma) = \Big(z^{\mu}_A\, l^A\Big)(\tau, \vec
\sigma)$  to the 3-spaces as unit 4-velocity; the world-lines of
these observers are the integral curves of the unit normal and in
general are not geodesics; in adapted radar 4-coordinates  the
Eulerian observers carry the tetrads defined in Eqs.(\ref{4.1});
ii) The skew congruence with unit 4-velocity $v^{\mu}(\tau, \vec
\sigma) = \Big(z^{\mu}_A\, v^A\Big)(\tau, \vec \sigma)$ (in general
it is not surface-forming, i.e. it has a non-vanishing vorticity);
the observers of the skew congruence have the world-lines (integral
curves of the 4-velocity) defined by $\sigma^r = const.$ for every
$\tau$, because the unit 4-velocity tangent to the flux lines
$x^{\mu}_{{\vec \sigma}_o}(\tau) = z^{\mu}(\tau, {\vec \sigma}_o)$
is $v^{\mu}_{{\vec \sigma}_o}(\tau) = z^{\mu}_{\tau}(\tau, {\vec
\sigma}_o)/\sqrt{\sgn\, {}^4g_{\tau\tau}(\tau, {\vec \sigma}_o)}$;
see Ref.\cite{84} for their adapted tetrads and for their relevance in the description of physics in GR.

\subsection{Hamiltonian ADM Tetrad Gravity}

The 16 configurational variables in the ADM action are
$\varphi_{(a)}$, $1 + n$, $n_{(a)}$, ${}^3e_{(a)r}$.
Their conjugate momenta are $\pi_{\varphi}$, $\pi_n$, $\pi_{n_{(a)}}$,
${}^3\pi^r_{(a)}$. There are ten
primary constraints (the vanishing of the 7 momenta of boosts, lapse
and shift variables plus three constraints inducing the rotation
on the flat indices $(a)$ of the cotriads) and four secondary ones
(the super-Hamiltonian and super-momentum constraints): all of them
are first class in the phase space spanned by 16+16 fields. This
implies that there are 14 gauge variables describing {\it inertial
effects} and 2 canonical pairs of physical degrees of freedom
describing the {\it tidal effects} of the gravitational field
(namely gravitational waves in the weak field limit). In this
canonical basis only the momenta ${}^3\pi^r_{(a)}$ conjugated to the
cotriads are not vanishing. The basis of canonical variables for this formulation of tetrad
gravity, naturally adapted to 7 of the 14 first-class constraints,
is

\beq
 \begin{minipage}[t]{3cm}
\begin{tabular}{|l|l|l|l|} \hline
$\varphi_{(a)}$ & $n$ & $\bar n_{(a)}$ & ${}^3e_{(a)r}$ \\ \hline $
\pi_{\varphi_{(a)}}\, \approx 0$ & $\pi_n\, \approx 0$ &
$\pi_{n_{(a)}}\, \approx 0 $ & ${}^3{ \pi}^r_{(a)}$
\\ \hline
\end{tabular}
\end{minipage}
 \label{4.3}
 \eeq

From Eqs.(5.5) of  Ref.\cite{78} we assume the
following (direction-independent, so to kill super-translations)
boundary conditions at spatial infinity ($r = \sqrt{\sum_r\,
(\sigma^r)^2}$; $\epsilon > 0$; $M = const.$): $n(\tau, \sigma^r)
\rightarrow_{r\, \rightarrow\, \infty}\, O(r^{- (2 + \epsilon )})$,
$\pi_n(\tau, \sigma^r) \rightarrow_{r\, \rightarrow\, \infty}\,
O(r^{-3})$, $n_{(a)}(\tau, \sigma^r) \rightarrow_{r\, \rightarrow\,
\infty}\, O(r^{- \epsilon })$, $\pi_{n_{(a)}}(\tau, \sigma^r)
\rightarrow_{r\, \rightarrow\, \infty}\, O(r^{-3})$,
$\varphi_{(a)}(\tau, \sigma^r) \rightarrow_{r\, \rightarrow\,
\infty}\, O(r^{- (1 + \epsilon)})$, $\pi_{\varphi_{(a)}}(\tau,
\sigma^r) \rightarrow_{r\, \rightarrow\, \infty}\, O(r^{-2})$,
${}^3e_{(a)r}(\tau, \sigma^r) \rightarrow_{r\, \rightarrow\,
\infty}\, \Big(1 + {M\over {2 r}}\Big)\, \delta_{ar} + O(r^{-
3/2})$, ${}^3\pi^r_{(a)}(\tau, \sigma^r) \rightarrow_{r\,
\rightarrow\, \infty}\, O(r^{- 5/2})$.

\subsection{The York Canonical Basis: Inertial Gauge Variables and Tidal Physical Degrees of Freedom}

In Ref.\cite{81}  a partial Shanmugadhasan canonical transformation to a
canonical basis adapted to ten of the first class constraints was found. It
implements the York map of Ref.\cite{86} (in the cases in which the
3-metric ${}^3g_{rs}$ has three distinct eigenvalues) and
diagonalizes the York-Lichnerowicz approach. Its final form is ($\alpha_{(a)}(\tau, \sigma^r)$ are the
Euler angles of the previous Subsection;  $V_{ua} {\buildrel {def}\over =} \sum_v\, V_{uv}\, \delta_{v(a)}$)

\bea
 &&\begin{minipage}[t]{4 cm}
\begin{tabular}{|ll|ll|l|l|l|} \hline
$\varphi_{(a)}$ & $\alpha_{(a)}$ & $n$ & ${\bar n}_{(a)}$ &
$\theta^r$ & $\tilde \phi$ & $R_{\bar a}$\\ \hline
$\pi_{\varphi_{(a)}} \approx0$ &
 $\pi^{(\alpha)}_{(a)} \approx 0$ & $\pi_n \approx 0$ & $\pi_{{\bar n}_{(a)}} \approx 0$
& $\pi^{(\theta )}_r$ & $\pi_{\tilde \phi} = {{c^3}\over {12\pi G}}\, {}^3K$ & $\Pi_{\bar a}$ \\
\hline
\end{tabular}
\end{minipage}\nonumber \\
 &&{}\nonumber \\
 &&{}\nonumber \\
 &&{}^3e_{(a)r} = \sum_b\, R_{(a)(b)}(\alpha_{(c)})\, {}^3{\bar e}_{(b)r},\qquad
 {}^3{\bar e}_{(a)r} = V_{ra}(\theta^i)\,
 {\tilde \phi}^{1/3}\, e^{\sum_{\bar a}^{1,2}\, \gamma_{\bar aa}\, R_{\bar a}},\nonumber \\
 &&{}\nonumber \\
 &&{}^4g_{\tau\tau} = \sgn\, [(1 + n)^2 - \sum_a\, {\bar n}^2_{(a)}],
 \qquad {}^4g_{\tau r} = - \sgn\, \sum_a\, {\bar
 n}_{(a)}\, {}^3{\bar e}_{(a)r},\nonumber \\
 &&{}^4g_{rs} = - \sgn\, {}^3g_{rs} = - \sgn\, {\tilde \phi}^{2/3}\,
 \sum_a\, V_{ra}(\theta^i)\, V_{sa}(\theta^i)\, Q^2_a,\qquad
 Q_a = e^{ \sum_{\bar a}^{1,2}\, \gamma_{\bar aa}\, R_{\bar
 a}},\nonumber \\
 &&{}\nonumber \\
 \label{4.4}
 \eea

The set of numerical parameters $\gamma_{\bar aa}$ satisfies
\cite{75} $\sum_u\, \gamma_{\bar au} = 0$, $\sum_u\, \gamma_{\bar a
u}\, \gamma_{\bar b u} = \delta_{\bar a\bar b}$, $\sum_{\bar a}\,
\gamma_{\bar au}\, \gamma_{\bar av} = \delta_{uv} - {1\over 3}$.
Each solution of these equations defines a different York canonical
basis. This canonical basis has been found  due to the fact that the
3-metric $ {}^3g_{rs}$ is a real symmetric $3 \times 3$ matrix,
which may be diagonalized with an {\it orthogonal} matrix
$V(\theta^r)$, $V^{-1} = V^T$, $det\, V = 1$, depending
on three parameters $\theta^r$. If we choose these three gauge
parameters to be Euler angles ${\hat \theta}^i(\tau, \vec \sigma)$,
we get a description of the 3-coordinate systems on $\Sigma_{\tau}$
from a local point of view, because they give the orientation of the
tangents to the three 3-coordinate lines through each point.
However, for the calculations (see Refs.\cite{82}) it is more
convenient to choose the three gauge parameters as first kind
coordinates $\theta^i(\tau, \vec \sigma)$ ($- \infty < \theta^i < +
\infty$) on the O(3) group manifold.  From now on for the sake of notational simplicity we shall use  $V$ for
$V(\theta^i)$.

\bigskip

This canonical transformation realizes a {\it York map} because the
gauge variable $\pi_{\tilde \phi}$  is proportional to {\it York internal
extrinsic time} ${}^3K$. It is the only gauge variable among the
momenta: this is a reflex of the Lorentz signature of space-time,
because $\pi_{\tilde \phi}$ and $\theta^n$ can be used as a set of
4-coordinates \cite{87}. The York time describes the effect of
gauge transformations producing a deformation of the shape of the
3-space along the 4-normal to the 3-space as a 3-sub-manifold of
space-time.
Its conjugate variable, to be determined by the super-Hamiltonian
constraint, is $\tilde \phi  = {}^3\bar e = \sqrt{det\,
{}^3g_{rs}}$, which is proportional to {\it Misner's internal
intrinsic time}; moreover $\tilde \phi$ is the {\it 3-volume
density} on $\Sigma_{\tau}$: $V_R = \int_R d^3\sigma\, \tilde \phi$,
$R \subset \Sigma_{\tau}$. Since we have ${}^3g_{rs} = {\tilde
\phi}^{2/3}\, {}^3{\hat g}_{rs}$ with $det\, {}^3{\hat g}_{rs} = 1$,
$\tilde \phi$ is also called the {\it conformal factor} of the
3-metric.

\medskip

The two pairs of canonical variables $R_{\bar a}$, $\Pi_{\bar a}$,
$\bar a = 1,2$, describe the generalized {\it tidal effects}, namely
the independent physical degrees of freedom of the gravitational
field. They are 3-scalars on $\Sigma_{\tau}$ and
the configuration tidal variables $R_{\bar a}$ depend {\it only on
the eigenvalues of the 3-metric}. They are DO's {\it
only} with respect to the gauge transformations generated by 10 of
the 14 first class constraints.

 \medskip

Since the variables $\tilde \phi$  and $\pi_i^{(\theta )}$
are determined by the super-Hamiltonian (i.e.
the Lichnerowicz equation) and super-momentum constraints respectively
(they are coupled elliptic PDE's in the 3-space $\Sigma_{\tau}$), the {\it
arbitrary gauge variables} are $\alpha_{(a)}$, $\varphi_{(a)}$,
$\theta^i$, $\pi_{\tilde \phi}$, $n$ and ${\bar n}_{(a)}$. As shown
in Refs.\cite{75,78,82}, they describe the following generalized {\it
inertial effects}:

\noindent a) $\alpha_{(a)}(\tau ,\vec \sigma )$ and $\varphi_{(a)}(\tau ,\vec
\sigma )$ are the 6 configuration variables parametrizing the O(3,1)
gauge freedom in the choice of the tetrads in the tangent plane to
each point of $\Sigma_{\tau}$ and describe the arbitrariness in the
choice of a tetrad to be associated to a time-like observer, whose
world-line goes through the point $(\tau ,\vec \sigma )$; they fix
{\it the unit 4-velocity of the observer and the conventions for the
orientation of three gyroscopes and their transport along the
world-line of the observer}; the  {\it Schwinger time gauges} are
defined by the gauge fixings $\alpha_{(a)}(\tau, \vec \sigma)
\approx 0$, $\varphi_{(a)}(\tau, \vec \sigma) \approx 0$.

\noindent b) $\theta^i(\tau ,\vec \sigma )$
describe the arbitrariness in the choice of the 3-coordinates in the
instantaneous 3-spaces $\Sigma_{\tau}$ of the chosen non-inertial
frame  centered on an arbitrary time-like observer; their choice
will induce a pattern of {\it relativistic inertial forces} for the
gravitational field, whose potentials are the functions
$V_{ra}(\theta^i)$ present in the weak ADM energy ${\hat E}_{ADM}$.

\noindent c) ${\bar n}_{(a)}(\tau ,\vec \sigma )$, the shift functions,
describe which points on different instantaneous 3-spaces have the
same numerical value of the 3-coordinates; they are the inertial
potentials describing the effects of the non-vanishing off-diagonal
components ${}^4g_{\tau r}(\tau ,\vec \sigma )$ of the 4-metric,
namely they are the {\it gravito-magnetic potentials} \footnote{In
the post-Newtonian approximation in harmonic gauges they are the
counterpart of the electro-magnetic vector potentials describing
magnetic fields \cite{88}: A) $N = 1 + n$, $n\, {\buildrel
{def}\over =}\, - {{4\, \sgn}\over {c^2}}\, \Phi_G$ with $\Phi_G$
the {\it gravito-electric potential}; B) $n_r\, {\buildrel
{def}\over =}\, {{2\, \sgn}\over {c^2}}\, A_{G\, r}$ with $A_{G\,
r}$ the {\it gravito-magnetic} potential; C) $E_{G\, r} =
\partial_r\, \Phi_G - \partial_{\tau}\, ({1\over 2}\, A_{G\, r})$ (the {\it
gravito-electric field}) and $B_{G\, r} = \epsilon_{ruv}\,
\partial_u\, A_{G\, v} = c\, \Omega_{G\, r}$ (the {\it
gravito-magnetic field}). Let us remark that in arbitrary gauges the
analogy with electro-magnetism  breaks down.} responsible of effects
like the dragging of inertial frames (Lens-Thirring effect)
in the post-Newtonian approximation; the shift functions
are determined by the $\tau$-preservation of the gauge fixings
determining the gauge variables $\theta^i(\tau, \vec \sigma)$ \cite{58}.

\noindent d) $\pi_{\tilde \phi}(\tau ,\vec \sigma )$, i.e. the York time
${}^3K(\tau ,\vec \sigma )$, describes the non-dynamical
arbitrariness in the choice of the convention for the
synchronization of distant clocks which remains in the transition
from SR to GR; since the York time is present in the Dirac
Hamiltonian, it is a {\it new inertial potential} connected to the
problem of the relativistic freedom in the choice of the {\it shape
of the instantaneous 3-space}, which has no Newtonian analogue (in
Galilei space-time time is absolute and there is an absolute notion
of Euclidean 3-space); its effects are completely unexplored.

\noindent e) $1 + n(\tau ,\vec \sigma )$, the lapse function appearing in the
Dirac Hamiltonian, describes the arbitrariness in the choice of the
unit of proper time in each point of the simultaneity surfaces
$\Sigma_{\tau}$, namely how these surfaces are packed in the 3+1
splitting; the lapse function is determined by the
$\tau$-preservation of the gauge fixing for the gauge variable
${}^3K(\tau, \vec \sigma)$ \cite{58}.

\bigskip

See the first paper in Refs.\cite{82} for the expression of the
super-momentum constraints  ${\cal H}_{(a)}(\tau, \vec \sigma)
\approx 0$ [Eqs.(3.41)-(3.42)] and of the super-Hamiltonian
constraint ${\cal H}(\tau, \vec \sigma) \approx 0$ (the Lichnerowicz
equation) [Eqs.(3.44)-(3.45)]. The weak ADM energy is given in Eqs.
(3.43)-(3.45) of that paper, while the other weak Poincar\'e
generators are given in Eqs.(3.47). The expression of the
weak ADM energy in terms of the expansion ($\theta = - \sgn\, {}^3K
= - \sgn\, {{12\pi G}\over {c^3}}\, \pi_{\tilde \phi}$) and shear
$\sigma_{(a)(b)} = \sigma_{(b)(a)} =
({}^3K_{rs} - {1\over 3}\, {}^3g_{rs}\, {}^3K)\, {}^3{\bar
e}^r_{(a)}\, {}^3{\bar e}^s_{(b)}$ with $ \sum_a\, \sigma_{(a)(a)} =
0$  ( so that ${}^3K_{rs} = - {{\sgn}\over 3}\, {}^3g_{rs}\, \theta +
 \sigma_{(a)(b)}\, {}^3{\bar e}_{(a)r}\, {}^3{\bar e}_{(b)s}$)
of the Eulerian observers (see the first paper in Ref.\cite{82})) is

 \bea
 {\hat E}_{ADM} &=& c\, \int d^3\sigma\, \Big[{\check {\cal M}} -
  {{c^3}\over {16\pi\, G}}\, {\cal S} + {{4\pi\, G}\over {c^3}}\,
  {\tilde \phi}^{-1}\, \sum_{\bar b}\, \Pi^2_{\bar b} +\nonumber \\
  &+& \tilde \phi\, \Big( {{c^3}\over {16\pi\, G}}\,
 \sum_{ab, a\not= b}\, \sigma^2_{(a)(b)}
 - {{6\pi\, G}\over {c^3}}\, \pi^2_{\tilde \phi}\Big)
 \,\, \Big](\tau ,\vec \sigma ),
 \label{4.5}
 \eea

\noindent where ${\check {\cal M}} = \tilde \phi\, (1 + n)^2\,
T^{\tau\tau}$ is the energy-mass density of the matter (with
energy-momentum tensor $T^{AB}$) and ${\cal S}(\tilde \phi,
\theta^i, R_{\bar a})$ is an inertial potential depending on the
choice of the 3-coordinates in the 3-space (it is the
$\Gamma-\Gamma$ term in the scalar 3-curvature of the 3-space).
In $ {\hat E}_{ADM}$ there is a negative
kinetic term proportional to $({}^3K)^2$, vanishing only in the gauges ${}^3K(\tau, \vec \sigma) =
0$, which comes from the term bilinear in momenta present both in the
super-Hamiltonian and in the weak ADM energy: it was known that this
quadratic form was not definite positive but only in the York
canonical basis this can be made explicit.

\medskip

Finally the Dirac Hamiltonian is
$H_D = {1\over c}\, {\hat E}_{ADM} + \int d^3\sigma\, \Big[ n\,
{\cal H} - {\bar n}_{(a)}\, {\cal H}_{(a)}\Big](\tau ,\vec \sigma )
+ \lambda_r(\tau )\, {\hat P}^r_{ADM} +
 \int d^3\sigma\, \Big[\lambda_n\, \pi_n + \lambda_{
{\bar n}_{(a)}}\, \pi_{{\bar n}_{(a)}} + \lambda_{\varphi_{(a)}}\,
\pi_{ \varphi_{(a)}} + \lambda_{\alpha_(a)}\,
\pi^{(\alpha)}_{(a)}\Big](\tau ,\vec \sigma )$,
where the $\lambda_{...}(\tau, \vec \sigma)$'s are Dirac
multipliers. In particular the Dirac multiplier $\lambda_r(\tau)$
implements the rest frame condition ${\hat P}^r_{ADM} \approx 0$
required by the absence of super-translations.

\medskip

Once a gauge is completely fixed by giving the six gauge-fixings for
the O(3,1) variables $\varphi_{(a)}$, $\alpha_{(a)}$ (choice of the
tetrads and of their transport) and four gauge-fixings for
$\theta^i$ (choice of the 3-coordinates on the 3-space) and ${}^3K$
(determination of the shape of the 3-space as a 3-sub-manifold of
space-time by means of a clock synchronization convention), the
Hamilton equations generated by
the Dirac Hamiltonian  (replacing the standard 12 ADM equations and
the matter equations ${}^4\nabla_A\, T^{AB} = 0$)
become a deterministic set of coupled PDE's for
the lapse and shift functions (primary inertial gauge variables
\footnote{As said in Ref.\cite{58} their gauge fixings are induced by
putting the gauge fixings for the secondary gauge variables $\theta^i$
and ${}^3K$ into their Hamilton equations.
Among the Hamilton equations there are
the contracted Bianchi identities, namely the evolution equations
for the solutions $\tilde \phi$ and $\pi_i^{(\theta)}$ of the
constraints (they say that given a solution of the constraints on a
Cauchy surface, it remains a solution also at later times).
Instead in numerical gravity one gives
independent gauge fixings for both the primary and secondary gauge
variables in such a way to minimize the computer time.}), the tidal
variables and the matter. Given a solution $\pi_i^{(\theta)}$ and $\tilde \phi$ of
the super-momentum \footnote{See Refs.\cite{80,81} for the  generalized Gribov
ambiguity in metric and tetrad gravity arising in the solution of
the super-momentum constraints after having done the York map.} and
super-Hamiltonian constraints and the Cauchy data for the tidal
variables on an initial 3-space, we can find a solution for the tidal
variables $R_{\bar a}$, $\Pi_{\bar a}$, of their hyperbolic evolution Hamilton equations and
therefore a solution of Einstein's equations in radar 4-coordinates adapted to a time-like
observer in the chosen gauge.
\bigskip

In Refs.\cite{83} there is the Hamiltonian expression of radar tensors which
coincide with the Riemann and Weyl tensors on the
solutions of Einstein equations. Moreover, by using the time-like normal to the
3-spaces and the space-like direction identified by the shift function (for
solutions of Einstein equations in which it is not identically zero) it is possible
to build null tetrads and to give the Hamiltonian formulation of the Newman-Penrose
formalism \cite{85}. Therefore we get the Hamiltonian expression of Ricci and Weyl
scalars and of the eigenvalues of the Weyl tensor. It is shown that the Bergmann
observables \cite{89,90},  built with these eigenvalues cannot be DO's \cite{87}.

\subsection{The Search of Dirac Observables and Comments on the Canonical Quantization of Gravity}

As already said the tidal variables $R_{\bar a}$, $\Pi_{\bar a}$, are DO's {\it
only} with respect to the gauge transformations generated by 10 of
the 14 first class constraints. Let us remark that, if we fix
completely the gauge and we go to Dirac brackets, then the only
surviving dynamical variables $R_{\bar a}$ and $\Pi_{\bar a}$ become
two pairs of {\it non canonical} DO's for that gauge:
the two pairs of canonical DO's have to be found as a
Darboux basis of the copy of the reduced phase space identified by
the gauge and they will be (in general non-local) functionals of the
$R_{\bar a}$, $\Pi_{\bar a}$ variables.
Till now there exist only statements about the existence of global DO's for gravity \cite{91,92,93},
but no determination of them.
\bigskip

As shown in the second paper of Ref. \cite{83},the DO's of GR could be found with a
Shanmugadhasan canonical transformation adapted to all the constraints if a
global solution $\tilde \phi \approx f$, $\pi_i^{(\theta)} \approx f_i$
of the super-Hamiltonian and super-momentum constraints would be known.
In the final canonical basis there would be the replacement of the variables $\tilde \phi$ and $\pi_i^{(\theta)}$
with the Abelian constraints $\hat \phi = \tilde \phi - f \approx 0$ and ${\hat \pi}_i^{(\theta)} =
\pi_i^{(\theta)} - f_i \approx 0$. The global DO's would be the tidal variables ${\hat R}_{\bar a}$, ${\hat \Pi}_{\bar a}$,
of this canonical basis, whose inertial gauge variables would be ${\hat \pi}_{\tilde \phi}$ and ${\hat \theta}^i$
and the old lapse and shift functions.

\bigskip

Let us add a comment on the open problem of the quantization of gravity. In most of the
traditional approaches one quantizes all the
components of the 4-metric, namely both inertial and tidal variables,
and one is facing every type of problems from the interpretation of the wave function of the universe
in the Wheeler-DeWitt approach to the disappearing of space-time in the loop quantum gravity approach.

The finding of global DO's (or at least of approximate ones after a
HPM linearization like in the second paper of Ref.\cite{83})
would allow to quantize only these tidal physical variables
{\it leaving the inertial gauge variables as c-numbers}
(the multi-temporal quantization proposed in Ref. \cite{94} and studied in Refs.\cite{33,34}).
The space-time manifold, its 3-1 splittings and the inertial gauge variables (lapse, shift, 3-coordinates, final York time)
would remain classical notions determined by relativistic metrology:
only the eigenvalues of the 3-metric of the non-Euclidean 3-spaces would be quantized with an
induced quantization of 3-lengths, 3-areas and 3-volumes to be compared with the results of loop quantum gravity
(the study of Ashtekar variables \cite{9,95} in asymptotically Minkowskian space-times in the York
canonical basis is still to be done).

\subsection{The Einstein-Electromagnetism-Particle System and the Problem of Dark Matter}

In the three papers of Ref.\cite{82} there is the study of charged positive-energy point particles plus the electro-magnetic field
in tetrad gravity. In the second paper there is the definition of the HPM approximation
based on the fact that in the allowed 3+1 splittings of the asymptotically Minkowskian space-times
the 3-spaces are asymptotically Euclidean and there is an asymptotic Minkowski 4-metric.
Then one can do the Post-Newtonian (PN) approximation of the HPM one.
The natural family of gauges to be used with the York canonical basis are the
{\it non-harmonic  3-orthogonal Schwinger time gauges} (in them the 3-metric of the 3-spaces is diagonal
due to the gauge fixings $\theta^i(\tau, \vec \sigma) \approx 0$).
By using Grassmann regularization of both the electro-magnetic and gravitational self-energies,
we are able to recover the HPM description of gravitational waves (GW) (see Ref.\cite{96}) from binaries
like in Damour-Deruelle approach \cite{97}. Also relativistic fluids have studied in this framework \cite{47}.
In Subsection IIIB of the second paper in Refs.\cite{82} it is
shown that this HPM linearization can be interpreted as the first
term of a Hamiltonian PM expansion in powers of the Newton constant
$G$ in the family of 3-orthogonal gauges. This expansion has still
to be studied.

\medskip

In the case of positive-energy relativistic particles in absence of the electro-magnetic field
one can show that the HPM effective force acting on them contains:
a) the contribution of the lapse function ${\check n}_{(1)}$, which
generalizes the Newton force;
b) the contribution of the shift functions ${\check {\bar
n}}_{(1)(r)}$, which gives the gravito-magnetic effects;
c) the retarded contribution of GW's;
d) the contribution of the volume element $\phi_{(1)}$ (${\tilde
\phi} = 1 + 6\, \phi_{(1)} + O(\zeta^2)$), always summed to the
GW's, giving forces of Newton type;
e) the contribution of the inertial gauge variable (the non-local
York time) ${}^3{\cal K}_{(1)} = {1\over {\triangle}}\,
{}^3K_{(1)}$. Then one does the PN expansion with the following
final form of PN equations of motion for the particle $i$ with mass $m_i$

 \bea
 \frac{d}{dt}\Big[ \,m_i\Big(1+\frac{1}{c}\,\frac{d}{dt}\,{}^3{\tilde {\cal K}}_{(1)}(t,
 {\vec {\tilde \eta}}_i(t))\,\Big)\, {{d\, {\tilde \eta}_i^r(t)}\over {dt}}\Big]
 &\cir &\,-G\, {{\partial}\over {\partial\,
 {\tilde \eta}_i^r}}\,  \sum_{j \not= i}\, \eta_j\, {{m_i\,m_j}\over
 {|{\vec {\tilde \eta}}_i(t) - {\vec {\tilde \eta}}_j(t)|}}+{\cal O}(\zeta^2).
 \nonumber \\
 &&{}
 \label{4.6}
\eea

We see that the term in the non-local York time can be {\it
interpreted} as the introduction of an {\it effective (time-,
velocity- and position-dependent) inertial mass term} for the
kinetic energy of each particle: $m_i\, \mapsto\,
m_i\,\Big(1+\frac{1}{c}\,\frac{d}{dt}\,{}^3{\tilde {\cal
K}}_{(1)}(t, {\vec {\tilde \eta}}_i(t))\,\Big) = m_i + \triangle m_i$ in each
instantaneous 3-space. Instead in the Newton potential there are the
{\it gravitational masses} of the particles, equal to the inertial ones in
the 4-dimensional space-time due to the equivalence principle.
Therefore the effect is due to a modification of the effective
inertial mass in each non-Euclidean 3-space depending on its shape
as a 3-sub-manifold of space-time: {\it it is the equality of the
inertial and gravitational masses of Newtonian gravity to be
violated}! In Galilei space-time the Euclidean 3-space is an
absolute time-independent notion like Newtonian time: the
non-relativistic non-inertial frames live in this absolute 3-space
differently from what happens in SR and GR, where they are (in
general non-Euclidean) 3-sub-manifolds of the space-time \cite{34}.

\medskip

Now the {\it rotation curves of spiral galaxies} (see Refs.\cite{99,100}
for  reviews) imply that the relative 3-velocity of particles goes to constant
for large $r$ (instead of vanishing like in Kepler theory).This
result can be simulated by fitting $\triangle\, m(r)$ (i.e. the
non-local York time in Eq.(\ref{4.6})) to the experimental data:  as a consequence
$\triangle\, m(r)$ is interpreted as a {\it dark matter halo} around
the galaxy. With our approach this dark matter would be a {\it
relativistic inertial effect} consequence of the a non-trivial shape
of the non-Euclidean 3-space as a 3-sub-manifold of space-time.
A similar interpretation can be given for the other two  main
signatures of the existence of dark matter in the observed masses of
galaxies and clusters of galaxies, namely the virial theorem
\cite{101,102,103} and weak gravitational lensing
\cite{102,104}.
Therefore at least part of dark matter is a relativistic inertial effect
connected with the inertial gauge variable York time to be
solved with relativistic metrology \cite{16} finding a convenient metrology convention
for the fixation of York time. GAIA \cite{105} will give the lacking information for extending the
IAU relativistic conventions \cite{106} inside the solar system to the Milky Way (where till now
we have only non-relativistic conventions like in the extragalactic cosmological regions).

\section{Conclusions and Open Problems}

As I have shown all the relevant systems in SR and GR require Dirac
theory of constraints for their Hamiltonian formulation. When the constraints can be solved,
Shanmugadhasan canonical transformations adapted to all the
existing first- and second-class constraints allow to find the DO's, i.e. the global
physical degrees of freedom of the given system.

One big open problem is whether one must quantize only the DO's (when known) or also the gauge variables (un-physical
degrees of freedom in gauge theories and inertial variables connected with relativistic metrology in GR)
with a determination of the DO's only at the quantum level like in the BRST approach. When both the possibilities are open, one faces the risk to obtain
unitarily inequivalent descriptions of the given system.

\bigskip

In the inertial frames of SR  the quantization of DO's can be faced with the
standard methods.  The main open problem is the quantization of
the scalar Klein-Gordon field in non-inertial frames, due to the
Torre and Varadarajan \cite{107} no-go theorem, according to which
in general the evolution from an initial space-like hyper-surface
to a final one is {\it  not unitary} in the Tomonaga-Schwinger
formulation of QFT. From the 3+1 point of view
there is evolution only among the leaves of an admissible
foliation and the possible way out from the theorem lies in the
determination of all the admissible 3+1 splittings of Minkowski
space-time satisfying the following requirements: i) existence of
an instantaneous Fock space on each simultaneity surface
$\Sigma_{\tau}$ (i.e. the $\Sigma_{\tau}$'s must admit a
generalized Fourier transform); ii) unitary equivalence of the
Fock spaces on $\Sigma_{\tau_1}$ and $\Sigma_{\tau_2}$ belonging
to the same foliation (the associated Bogoljubov transformation
must be Hilbert-Schmidt), so that the non-inertial Hamiltonian is
a Hermitean operator; iii) unitary gauge equivalence of the 3+1
splittings with the Hilbert-Schmidt property. The overcoming of
the no-go theorem would help also in QFT in
curved space-times and in condensed matter (here the non-unitarity
implies non-Hermitean Hamiltonians and negative energies).

\bigskip

In GR the described approach to tetrad gravity has to be completed by facing the following topics
before trying to quantize the linearized HPM theory :

1) Find the second order of the HPM expansion to see whether in PM
space-times there is the emergence of hereditary terms (see
Refs.\cite{96,108}) like the ones present in harmonic gauges. Like
in standard approaches (see the review in Appendix A of the second
paper in Refs.\cite{82}) regularization problems may arise at the
higher orders.

2) Study  the PM equations of motion of the transverse
electro-magnetic field trying to find Lienard-Wiechert-type
solutions (see Subsection VB of the second paper in
Refs.\cite{82}). Study astrophysical problems where the
electro-magnetic field is relevant.

3) The next big challenge after dark
matter is {\it dark energy} in cosmology \cite{109,110}. Even if in cosmology we
cannot use canonical gravity, in the first paper of Ref.\cite{82}
it is shown that the usual non-Hamiltonian 12 ADM equations can be
put in a form allowing to use the interpretations based on the York
canonical basis by means of the expansion and the shear of the
Eulerian observers.
Let us remark that in the Friedmann-Robertson-Walker (FRW)
cosmological solution the Killing symmetries connected with
homogeneity and isotropy imply ($\tau$ is the cosmic time, $a(\tau)$
the scale factor) ${}^3K (\tau) = - {{\dot a(\tau)}\over {a(\tau)}}
= - H$, namely the York time is no more a gauge variable but
coincides with the Hubble constant. However at the first order in
cosmological perturbations we have ${}^3K = - H + {}^3K_{(1)}$ with
${}^3K_{(1)}$ being again an inertial gauge variable. Instead in
inhomogeneous space-times without Killing symmetries like the
Szekeres ones \cite{111} the York time remains an inertial gauge
variable. Therefore the York time has a central position also in the main
quantities on which relies the interpretation of dark energy in the
standard $\Lambda$CDM cosmological model (Hubble constant, the old
Hubble redshift-distance relation  replaced in FRW cosmology with
the velocity distance relation or Hubble law). As a consequence it
looks reasonable to investigate on a possible gauge origin also of
dark energy.

\end{document}